\documentclass[fleqn,usenatbib]{mnras}

\usepackage{mathptmx}

\usepackage[T1]{fontenc}

\DeclareRobustCommand{\VAN}[3]{#2}
\let\VANthebibliography\thebibliography
\def\thebibliography{\DeclareRobustCommand{\VAN}[3]{##3}\VANthebibliography}

\usepackage{graphicx}	
\usepackage{amsmath}	
\usepackage{amssymb}	
\usepackage{xspace}
\usepackage{longtable}
\usepackage{subfig}
\usepackage{booktabs}
\usepackage{lscape}
\usepackage{siunitx}
\usepackage[symbol]{footmisc}
\usepackage[dvipsnames]{xcolor}

\newcommand{\tess}{\ensuremath{\emph{TESS}}\xspace}
\newcommand{\jwst}{\ensuremath{\emph{JWST}}\xspace}

\newcommand*{\swap}[2]{#2#1}

\title[A small habitable zone planet around TOI-715]{A 1.55~$\rm R_\oplus$ habitable-zone planet hosted by TOI-715, an M4 star near the ecliptic South Pole}

\author[G. Dransfield et al.]{
Georgina Dransfield,$^{1}$\thanks{E-mail: gxg831@bham.ac.uk}
Mathilde Timmermans,$^{2}$ 
Amaury H.M.J. Triaud,$^{1}$ 
Martín Dévora-Pajares,$^{3}$ 
\newauthor
Christian Aganze,$^{4}$ 
Khalid Barkaoui,$^{2,5,6}$ 
Adam J. Burgasser,$^{4}$ 
Karen A. Collins,$^{7}$ 
Marion Cointepas,$^{8,9}$ 
\newauthor
Elsa Ducrot,$^{10}$\thanks{Paris Region Fellow, Marie Sklodowska-Curie Action} 
Maximilian N. G{\"u}nther$^{11}$\thanks{ESA Research Fellow}, 
Steve B. Howell,$^{12}$ 
Catriona A. Murray,$^{13}$ 
Prajwal Niraula,$^{5}$ 
\newauthor
Benjamin V. Rackham,$^{5}$\thanks{51 Pegasi b Fellow} 
Daniel Sebastian,$^{1}$ 
Keivan G.\ Stassun$^{14}$, 
Sebasti\'an Z\'u$\rm \Tilde{n}$iga-Fern\'andez,$^{2}$ 
\newauthor
Jos\'e Manuel Almenara,$^{8}$ 
Xavier Bonfils,$^{8}$ 
Fran\c cois Bouchy,$^{9}$ 
Christopher~J.~Burke,$^{15}$ 
David~Charbonneau,$^{7}$ 
\newauthor
Jessie~L.~Christiansen,$^{16}$  
Laetitia Delrez,$^{2}$ 
Tianjun Gan,$^{18}$ 
Lionel J. Garc\'ia,$^{2}$ 
Micha\"el Gillon,$^{2}$ 
\newauthor
Yilen G\'omez Maqueo Chew,$^{19}$ 
Katharine~M.~Hesse,$^{15}$ 
Matthew J. Hooton,$^{20}$ 
Giovanni Isopi,$^{21}$
\newauthor
Emmanu\"el Jehin,$^{22}$ 
Jon~M.~Jenkins,$^{12}$ 
David~W.~Latham,$^{7}$ 
Franco Mallia,$^{21}$ 
Felipe Murgas,$^{8,6,23}$ 
\newauthor
Peter P. Pedersen,$^{20}$ 
Francisco J. Pozuelos,$^{2,22,24}$ 
Didier Queloz,$^{20}$
David~R.~Rodriguez,$^{25}$ 
\newauthor
Nicole Schanche,$^{17,26,27}$ 
Sara Seager,$^{4,23,26}$ 
Gregor Srdoc,$^{29}$ 
Chris Stockdale,$^{30}$ 
Joseph D. Twicken,$^{12,31}$ 
\newauthor
Roland~Vanderspek,$^{15}$
Robert Wells,$^{17}$ 
Joshua N.\ Winn,$^{32}$ 
Julien de Wit,$^{5}$ 
Aldo Zapparata,$^{21}$ 
\\
\\
{\it A list of affiliations is given at the end of the paper}
}
\date{Accepted 2023~May~10. Received 2023~May~09; in original form 2022~November~03}

\pubyear{2022}

\begin{document}
\label{firstpage}
\pagerange{\pageref{firstpage}--\pageref{lastpage}}
\maketitle

\begin{abstract}
A new generation of observatories is enabling detailed study of exoplanetary atmospheres and the diversity of alien climates, allowing us to seek evidence for extraterrestrial biological and geological processes. Now is therefore the time to identify the most unique planets to be characterised with these instruments. In this context, we report on the discovery and validation of TOI-715\,b, a $R_{\rm b}=1.55\pm 0.06\rm R_{\oplus}$ planet orbiting its nearby ($42$\,pc) M4 host (TOI-715/TIC 271971130) with a period  $P_{\rm b} = 19.288004_{-0.000024}^{+0.000027}$ days. TOI-715\,b was first identified by \tess and validated using ground-based photometry, high-resolution imaging and statistical validation. The planet's orbital period combined with the stellar effective temperature $T_{\rm eff}=3075\pm75~\rm K$ give this planet an instellation $S_{\rm b} = 0.67_{-0.20}^{+0.15}~\rm S_\oplus$, placing it within the most conservative definitions of the habitable zone for rocky planets. TOI-715\,b's radius falls exactly between two measured locations of the M-dwarf radius valley; characterising its mass and composition will help understand the true nature of the radius valley for low-mass stars. We demonstrate TOI-715\,b is amenable for characterisation using precise radial velocities and transmission spectroscopy. Additionally, we reveal a second candidate planet in the system, TIC~271971130.02, with a potential orbital period of $P_{02} = 25.60712_{-0.00036}^{+0.00031}$ days and a radius of $R_{02} = 1.066\pm0.092\,\rm R_{\oplus}$, just inside the outer boundary of the habitable zone, and near a 4:3 orbital period commensurability. Should this second planet be confirmed, it would represent the smallest habitable zone planet discovered by \tess to date.
\end{abstract}

\begin{keywords}
exoplanets -- planets and satellites: detection -- planets and satellites: terrestrial planets -- planets and satellites: fundamental parameters
\end{keywords}



\section{Introduction}


At long last the era of \jwst has arrived, and with it the age of detailed exoplanetary atmospheric characterisation \citep{jwstERS}. This achievement was unlocked just as the community hit another significant milestone: the discovery of the 5000$^{\rm {th}}$ planet beyond the solar system\footnote{Reported on 2022\,March\,21: \url{https://exoplanetarchive.ipac.caltech.edu/docs/exonews_archive.html}}. This ever-growing sample combined with the might of \jwst is now set to deepen our understanding of planets, including those in our solar system, on a sub-population level.

One sub-population of particularly enduring interest consists of small, potentially habitable planets orbiting cool M-type stars. With current instrumentation, M dwarfs represent our best hope of finding temperate terrestrial planets, the best example being the seven Earth-sized planets of the TRAPPIST-1 system orbiting an M8V type star \citep{trappist1}. The reduced radii of M dwarfs enable small planets to produce large transit depths; this combined with the shorter orbital periods of these more compact systems makes photometric monitoring from ground and space-based facilities considerably more feasible \citep[e.g.][]{Mearth,Carmenes,speculoos,Triaud2021}. In the context of atmospheric characterisation by transmission spectroscopy, bright, nearby M dwarfs are ideal planetary hosts as small temperate planets will transit frequently, enabling high signal-to-noise detections of atmospheric features with fewer hours of telescope time \citep{mdwarfop,Dressing2015,Morley2017}. Additionally, these low mass stars appear more likely to host small planets from a theoretical and observational point of view \citep{montgomery2009,dressing2013,bonfils2013,mulders2015,Alibert2017,he2017,Sabotta2021}.

The so-called `habitable zone' is a circumstellar region where an Earth-like planet could sustain liquid water on its surface \citep{1993Icar..101..108K}. With sometimes contradictory definitions of its boundaries, which depend on stellar spectral type, planetary albedo, mass and even cloud cover \citep{kopparapu2014,2003IJAsB...2..289U,2016ApJ...823....6R}, it can be challenging to categorise planets using this metric. However, the most widely applicable and conservative definition comes from \cite{2013ApJ...765..131K}, stating that rocky planets receiving between $0.42-0.842\,S_{\oplus}$ are in their star's HZ, irrespective of all other factors.

The activity of the M-dwarf host stars themselves is an additional factor to consider in the habitability of planets orbiting M dwarfs, as the effect of their frequent flaring is not fully known \citep{omalley2017}. Stellar flares could destroy a nascent atmosphere entirely \citep{proxcenflare,proxcenhab}, or they might provide the energy needed to catalyse biological processes \citep{Buccino2007,Patel2015,Lingam2017,Rimmer2018}. Additionally, we must look beyond our own terrestrial version of habitability, by exploring the possibility that biology might arise on water-worlds \citep{Madhu2021} or temperate sub-Neptunes \citep{Seager2021}, as both of these may have different HZ boundaries.

We owe much of our current understanding of exoplanetary demographics to the \textit{Kepler} mission \citep{Kepler}, and one of the most influential results to come out of statistical studies of the \textit{Kepler} sample was the bimodal radius distribution of sub-Neptune sized planets, with a gap between $1.5-2\,R_{\oplus}$ \citep{Fulton2017,vaneylen2018}. The current best explanations for this bimodality are core-powered mass loss \citep{lopez2013,Ginzburg2018,gupta2019}, photo-evaporation \citep{owen2013,Owen2017} and volatile-poor formation \citep{Lee2014,venturini2017}. While the exact location of this so-called `radius valley' depends on stellar mass among other things \citep{wu19,gupta2020}, it is still unclear whether it is present around M dwarfs or not \citep{cloutier2020}. Therefore, planets with sizes within the radius valley \citep[e.g.][]{cloutier2020b,cloutier2021,luque22a} help us to understand the shape and depth of this gap around M dwarfs. A recent study by \cite{Luque2022}, however, indicates that M-dwarf planets may have a density gap rather than a radius gap separating two populations of small planets (rocky and water worlds), and that these observations are also well explained by formation pathways that include disc-driven migration \citep{venturini202}.

At present \tess \citep{tess} is providing the community with ample new planets to improve our understanding of exoplanetary demographics, including by populating the habitable zone. Notably, while several `habitable zone' planets discovered by \tess have been confirmed \citep[e.g.][]{gilbert2020,vach2022}, none yet have fallen within the conservative habitable zone as described by \cite{2013ApJ...765..131K}---until now.

In this context we report on the discovery of TOI-715\,b, a small planet orbiting a nearby M4 star (TOI-715 / TIC 271971130). The planet's relatively long orbital period and cool host provide it with a mild instellation, placing TOI-715\,b comfortably within its star's conservative habitable zone. We additionally find a second candidate in the system that, if confirmed, could be \tess's smallest habitable zone planet discovered to date.

Our paper is organised as follows: we begin by characterising the host star using its spectral energy distribution and reconnaissance spectroscopy in Section \ref{sec:star}. We then describe the identification of planetary candidates in the data, first by \tess followed by our own search, in Section \ref{sec:sherlock}. In Section \ref{sec:followup} we describe our ground-based follow-up campaign and our procedures to validate the system, and in Section \ref{sec:analysis} we detail our global analysis of all available data. Finally, we contextualise our results in Section \ref{sec:discussion} and conclude in Section \ref{sec:conc}.

\section{Stellar Characterisation}
\label{sec:star}

TOI-715 (TIC 271971130) is a nearby \citep[$\rm 42\,pc$;][]{BJdist} M dwarf of spectral type M4. It is a high-proper-motion target with right ascension 07:35:24.56 hms and declination -73:34:38.67 dms (J2000, epoch 2015.5), placing it within \tess's continuous viewing zone (CVZ). 

As all our planetary information will be derived using the host star's parameters, we begin in the section that follows by characterising TOI-715. All photometry and stellar parameters adopted for this work can be found in Table \ref{tab:starpar}.

\begin{table}
\centering
\caption{Stellar parameters adopted for this work.}
\begin{tabular}{@{}lp{25mm}p{30mm}@{}}
\toprule
{\bf Designations} & \multicolumn{2}{p{65mm}}{TOI-715, TIC 271971130, 2MASS J07352425-7334388, APASS 33649915, Gaia DR2 5262666416118954368, UCAC4 083-012601, WISE J073524.46-733438.7} \\ \midrule
{\bf Parameter} & {\bf Value}              & {\bf Source} \\ \midrule
$\alpha$           & 07:35:24.56      & \cite{gaiaDR3cat} \\
$\delta$           & -73:34:38.67      & \cite{gaiaDR3cat} \\
Distance         & 42.46$\pm$0.03\,pc      & \cite{BJdist} \\
$\mu_{\alpha}$           & $\rm 82.67\pm0.02\,mas\,yr^{-1}$      & \cite{gaiaDR3cat} \\
$\mu_{\delta}$           & $\rm 9.92\pm0.02\,mas\,yr^{-1}$      & \cite{gaiaDR3cat} \\
$RV$           & $\rm +55.8\pm2.7\,km\,s^{-1}$      & \cite{gaiaDR3cat} \\
$U$           & $\rm +29.7\pm0.6\,km\,s^{-1}$      & This work \\
$V$           & $\rm -54.8\pm2.1\,km\,s^{-1}$      & This work \\
$W$           & $\rm +0.7\pm0.9\,km\,s^{-1}$      & This work \\
SpT             & M4                  & This work \\
$R_{\star}$     & $0.240\pm0.012R_{\odot}$ & This work                  \\
$M_{\star}$     & $0.225\pm0.012M_{\odot}$   & This work                  \\
${\rm T_{eff}}$ & 3075$\pm$75 K             & This work                  \\
$\log g_\star$           & 5.0$\pm$0.2           & This work                  \\
$\rm [Fe/H]$        & +0.09$\pm$0.20 dex              & This work (spectroscopy)                 \\   
    & $-$0.25$\pm$0.25 dex              & This work (SED)                 \\ 
 Age   & 6.2$^{+3.2}_{-2.2}$~Gyr              & This work                 \\ 
T mag           & 13.5308$\pm$0.0073        & \cite{TICv8} \\
B mag           & 18.14$\pm$0.16      & \cite{ucac4} \\
V mag           & 16.24$\pm$0.01        & \cite{ucac4} \\
G mag           & 14.8940$\pm$0.0007      & \cite{gaiaDR3cat} \\
J mag           & 11.808$\pm$0.024          & \cite{2masscat} \\
H mag           & 11.264$\pm$0.026          & \cite{2masscat} \\
K mag           & 10.917$\pm$0.019         & \cite{2masscat} \\
W1 mag           & 10.753$\pm$0.023          & \cite{wisecat} \\
W2 mag           & 10.571$\pm$0.020          & \cite{wisecat} \\
W3 mag           & 10.387$\pm$0.049            & \cite{wisecat} \\
W4 mag           & > 8.92       & \cite{wisecat} \\

\bottomrule
\end{tabular}
\label{tab:starpar}
\end{table}

\subsection{Reconnaissance Spectroscopy}
\label{sec:recspec}

We collected an optical spectrum of TOI-715 on 2022\,January\,07 (UT) using the Low Dispersion Survey Spectrograph on the 6.5-m \textit{Magellan} \textsc{II} (Clay) Telescope at Las Campanas Observatory in Chile.
With its upgraded red-sensitive CCD \citep{Stevenson2016}, this instrument is now known as ``LDSS-3C''.
We used LDSS-3C in long-slit mode with the standard setup (fast readout speed, low gain, and $1\times1$ binning) and the VPH-Red grism, OG-590 blocking filter, and the $0\farcs75 \times 0\farcm4$ center slit.
This setup provides spectra covering 6000--10\,000\,\AA\ with a resolution of $R \sim 1810$, which is insufficient to produce a notable constraint on $v\sin i_\star$.

We observed TOI-715 during clear conditions with seeing of $0\farcs5$.
We collected six exposures of 300\,s each, totaling 30\,min on-source, at an average airmass of 1.445.
Afterwards, we collected three 1-s exposures of the nearby F8\,V standard star HR\,2283 \citep{Maiolino1996} at an average airmass of 1.360.
At each pointing, we collected a 1-s HeNeAr arc lamp exposure and three, 10-s flat fields with the ``quartz high'' lamp.
We reduced the data with a custom, {\sc Python}-based pipeline, which includes bias removal, flat field correction, and spectral extraction. 
For wavelength calibration, the HeNeAr arc exposure was used, which was extracted similarly to the science spectrum.
For flux calibration, we used the ratio of the spectrum of the flux standard HR\,2283 with an F8\,V template from \citet{1998PASP..110..863P} to compute a relative flux correction; no correction was made to address telluric absorption. 
The final science spectrum has a maximum SNR per resolution element of 193 at 9202\,\AA\ and a mean SNR per resolution element of 124 in the 6000--10\,000\,\AA\ range.

\begin{figure}
    \centering
    \includegraphics[width =\columnwidth]{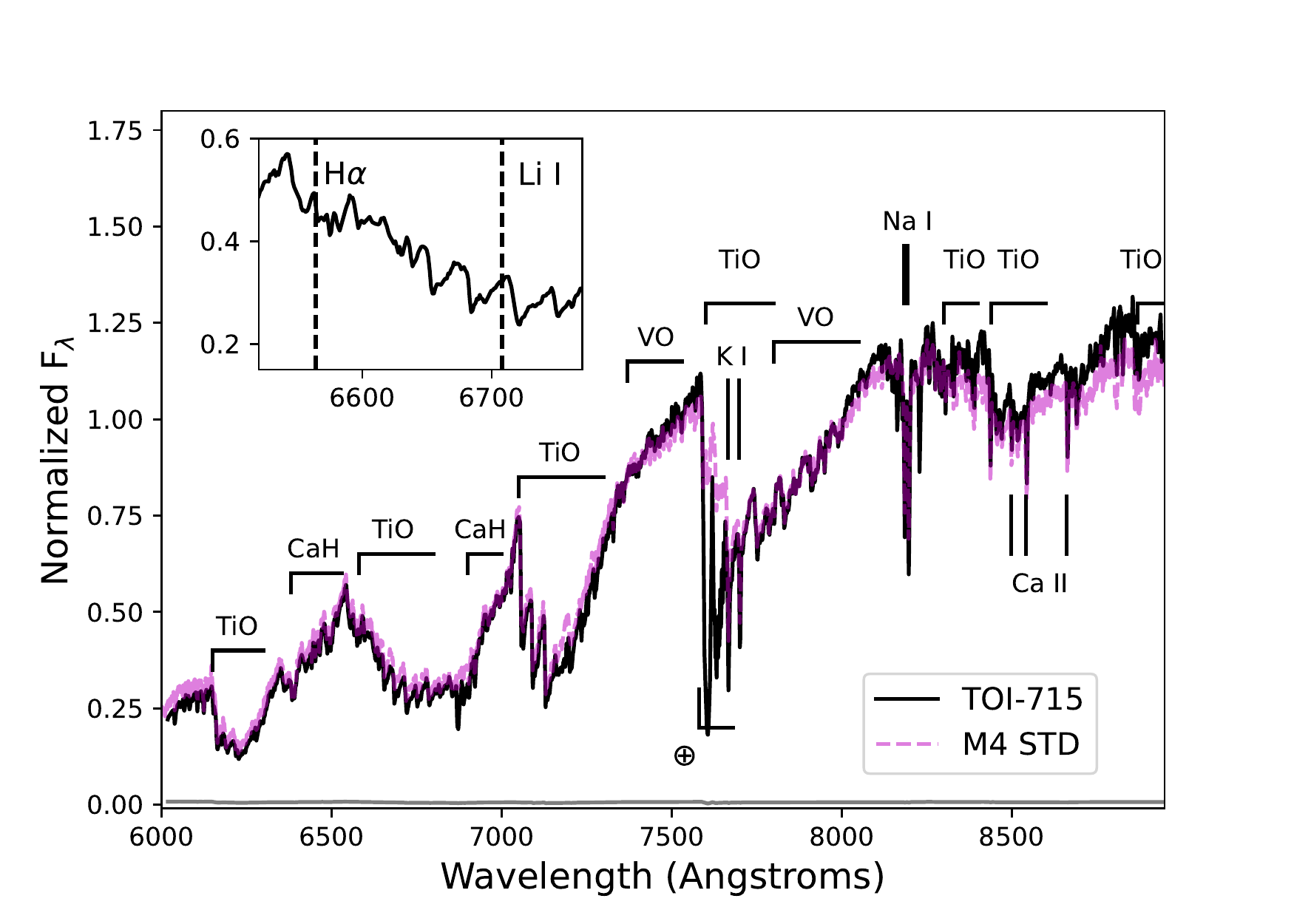}
    \caption{LDSS-3C red optical spectrum of TOI-715 (black line), compared to its best-fit M4 template \citep[magenta line]{2017ApJS..230...16K}. Spectra are normalized in the 7400--7500~{\AA} region, and major absorption features are labeled, including regions of strong telluric absorption ($\oplus$). The inset box shows a close-up of the region encompassing H$\alpha$ (6563~{\AA}) and Li\,\textsc{i}~ (6708~{\AA}) features, neither of which is detected in the data.  }
    \label{fig:ldss}
\end{figure}

The reduced spectrum (Figure~\ref{fig:ldss}) was analyzed using \textsc{kastredux}\footnote{\url{https://github.com/aburgasser/kastredux}.}.
We compared the spectrum to Sloan Digital Sky Survey templates from \citet{2017ApJS..230...16K}, finding a best match to an M4 dwarf. This classification was verified through spectral classification indices from \citet{1995AJ....110.1838R}; \citet{2003AJ....125.1598L}; and \citet{2007MNRAS.381.1067R}, all of which measure consistent spectral classifications of M4. We evaluated the $\zeta$ metallicity index \citep{2007ApJ...669.1235L,2013AJ....145..102L}, determining a value of 1.066$\pm$0.002 which corresponds to a metallicity [Fe/H] = $+$0.09$\pm$0.20\,dex based on the empirical calibration of \citet{2013AJ....145...52M}. There is no significant evidence of H$\alpha$ emission, with an equivalent width limit $|EW|$ $<$ 1.2~{\AA} corresponding to $\log_{10}{L_{H\alpha}/L_{bol}} < -6.5$ \citep{2014ApJ...795..161D}. We also find no evidence of Li~I absorption at 6708~{\AA} ($EW < 0.7$~{\AA}), ruling out a young age and substellar mass \citep{1993ApJ...404L..17M}.

\subsection{Spectral Energy Distribution}
\label{sec:sed}

\begin{figure}
    \centering
    \includegraphics[width =\columnwidth]{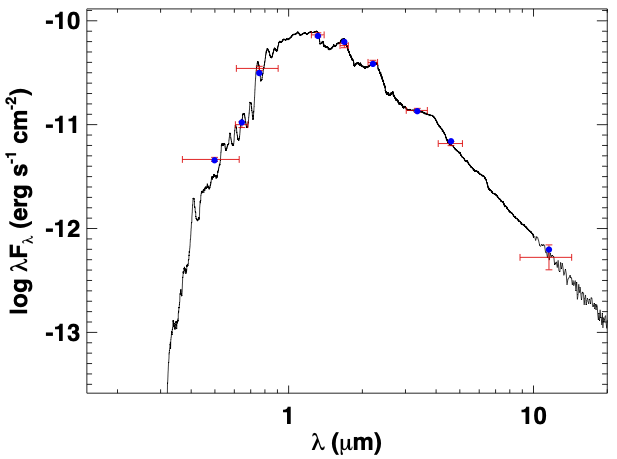}
    \caption{Spectral energy distribution of TOI-715. Red symbols represent the observed photometric measurements, where the horizontal bars represent the effective width of the passband. Blue symbols are the model fluxes from the best-fit NextGen atmosphere model (black). }
    \label{fig:SED}
\end{figure}

As an independent determination of the basic stellar parameters, we performed an analysis of the broadband spectral energy distribution (SED) of the star together with the {\it Gaia\/} DR3 parallax \citep[with no systematic offset applied; see, e.g.,][]{StassunTorres:2021}, to determine an empirical measurement of the stellar radius, following the procedures described in \citet{Stassun:2016} and \citet{Stassun:2017,Stassun:2018}. We utilised the highest quality broadband photometry available, which were the $JHK_S$ magnitudes from {\it 2MASS}, the W1--W3 magnitudes from {\it WISE}, and the $G G_{\rm BP} G_{\rm RP}$ magnitudes from {\it Gaia}. Together, the available photometry spans the full stellar SED over the wavelength range 0.4--10~$\mu$m (see Figure~\ref{fig:SED}).  

We performed a fit using NextGen stellar atmosphere models \citep{nextgen}, with the free parameters being the effective temperature ($T_{\rm eff}$), surface gravity ($\log g$), and metallicity ([Fe/H]). The remaining free parameter is the extinction $A_V$, which we fixed at zero due to the star's proximity. The resulting fit (Figure~\ref{fig:SED}) has a reduced $\chi^2$ of 1.3, with best-fit $T_{\rm eff} = 3075 \pm 75$~K, $\log g = 5.0 \pm 0.2$, [Fe/H] = $-0.25 \pm 0.25$. Integrating the (unreddened) model SED gives the bolometric flux at Earth, $F_{\rm bol} = 8.21 \pm 0.19 \times 10^{-11}$ erg~s$^{-1}$~cm$^{-2}$. Taking the $F_{\rm bol}$ and $T_{\rm eff}$ together with the {\it Gaia\/} parallax, gives the stellar radius, $R_\star = 0.240 \pm 0.012$~R$_\odot$. In addition, we can estimate the stellar mass from the empirical relations of \citet{Mann:2019}, giving $M_\star = 0.225 \pm 0.012$~M$_\odot$, which is consistent with the value of $0.21 \pm 0.05$~M$_\odot$ determined via $\log g$ and $R_\star$. 

\subsection{Estimated age}
\begin{figure}
    \centering
    \includegraphics[width=0.47\textwidth]{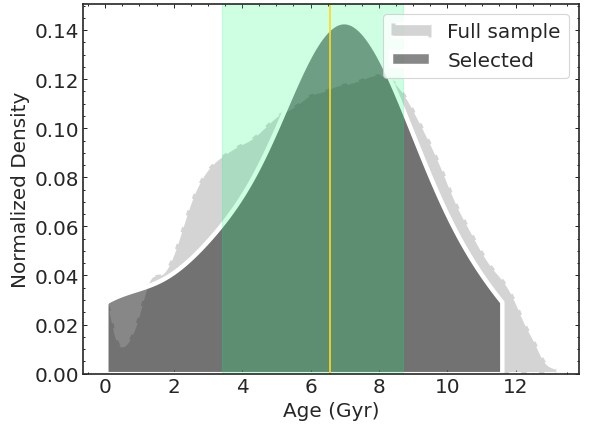}
    \caption{Distribution of ages for all sources in GALAH Data Release 3 (light grey histogram) and those GALAH sources that match the $UVW$ velocities and metallicity of TOI-715. (dark grey histogram). The latter distribution is consistent with a median age of 6.6$^{+3.2}_{-2.2}$~Gyr (green shaded region, 25\% and 75\% quantiles), which we adopt as the age of TOI-715.}
    \label{fig:age}
\end{figure}

The lack of detectable H$\alpha$ emission in its optical spectrum, and absence of any significant flaring activity in its TESS lightcurve, both suggest a relatively old age for TOI-715 given the persistent emission of most mid-type M dwarfs \citep{2017ApJ...834...85N,2019AJ....157..231K}.
We estimated the age of TOI-715 by comparing its $UVW$ velocities and metallicity to local stars with previously-determined ages from isochrone fitting and metallicities from high resolution spectroscopy (cf.~\citealt{2017ApJ...845..110B,speculoos2}). 
Our comparison sample was drawn from the GALAH Data Release 3 catalogue \citep{2021MNRAS.506..150B}, for which ages are estimated using the Bayesian Stellar Parameter Estimation ({\sc BSTEP}) code \citep{2018MNRAS.473.2004S}.
A matched GALAH sample was selected by requiring a distance $\leq$200~pc, and agreement with TOI-715 in individual $UVW$ velocities to within 10~km/s and metallicity to within 0.20~dex.
The age distributions of the full GALAH sample and matched stars are shown in Figure~\ref{fig:age}.
The latter shows a broad peak with a maximum probability at 7~Gyr; the median age of the distribution and 25\% and 75\% quantiles yields an age estimate of 6.6$^{+3.2}_{-2.2}$~Gyr.
TOI-715 is likely to be as old or older than the Sun, consistent with its low degree of magnetic activity. 

Analyses of ground-based and Kepler photometry by \cite{Newton2016} and \cite{McQuillan2014}, respectively, show that typical rotational periods of old M4 dwarfs ($0.4-0.5\,\rm M_\odot$) are roughly 15-50 days, most likely at the longer end of this range (and possibly beyond; longer periods in \cite{Newton2016} are "grade B" detections due to low amplitude variations). This range falls within the detectable period range for the full \tess dataset; however, the absence of detectable H$\alpha$ emission suggests a lack of significant spotting, so even if the period were within this range \tess measurements may not be sufficient to detect variability. In either case, the lack of H$\alpha$ emission and rotationally-induced variability on a time scale $\leq 50$ days are both consistent with an old age.

\section{Identification of planetary candidates}
\label{sec:sherlock}

\begin{figure*}
    \centering
    \includegraphics[width=0.95\textwidth]{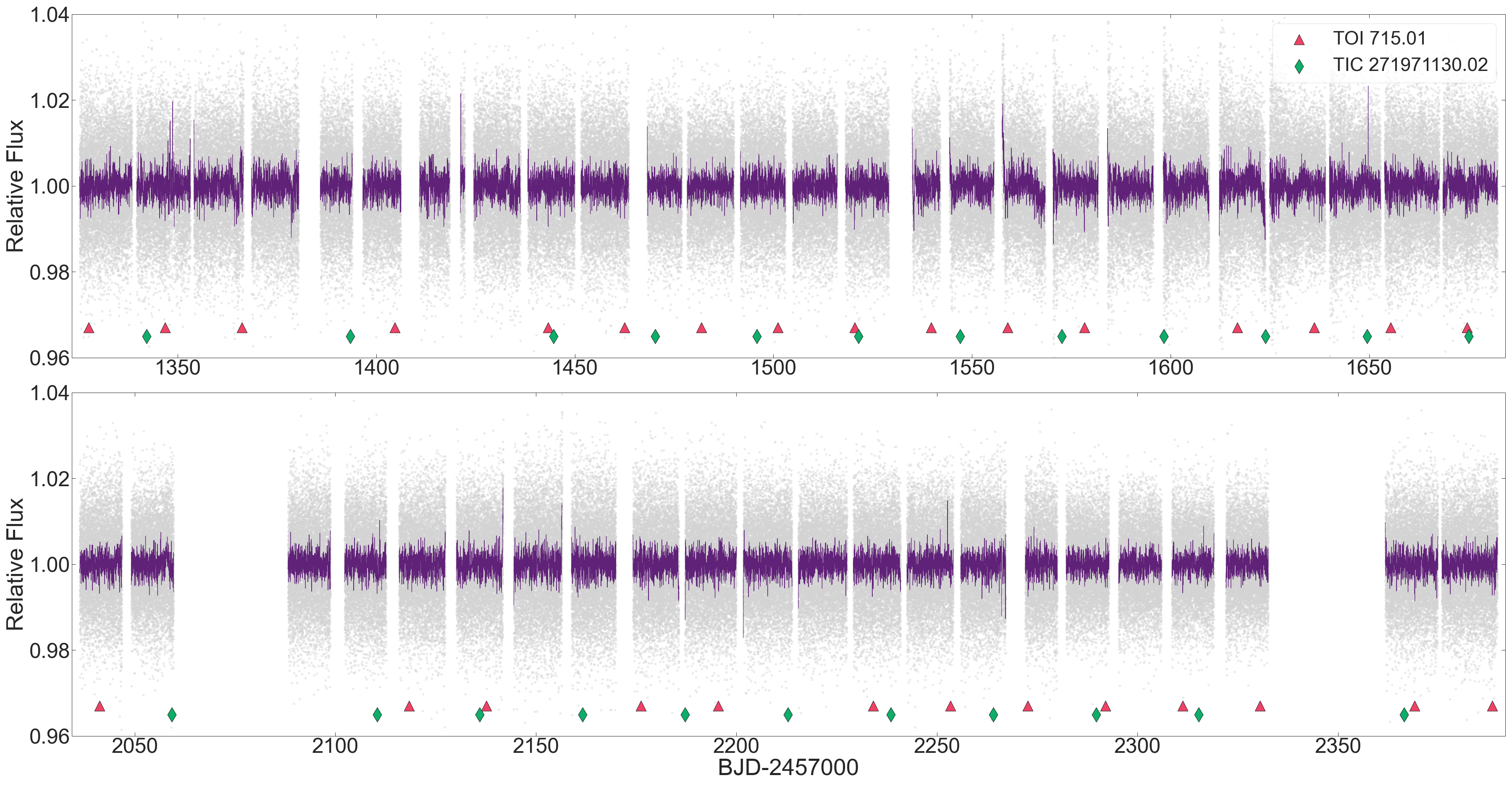}
    \caption{PDCSAP flux extracted from the short (2 minute) cadence data of the 24 sectors (1--13, 27, 29--37, 39) in which TOI-715 was observed. Light grey point show the $120\,\rm s$ exposures and the purple line shows the flux in $3\,\rm minute$ bins. The transit events of TOI-715.01 are shown with the dark pink arrows and the locations of possible transits of TIC~271971130.02 are indicated with green arrows; we note that individual transits are not readily apparent by eye.}
    \label{fig:tess}
\end{figure*}

TOI-715 is in \tess's southern continuous viewing zone (CVZ), meaning that in principle, it would be observed in all southern sectors. In practice, it was not optimally placed on the CCD during sectors 28 and 38, so at the time of writing, there are 24 sectors of short-cadence (2 minute) data for this target\footnote{There are additionally two sectors of fast cadence (20 second) data available for TIC 271971130 (27 and 28). Observations at 2 minute cadence were requested by two General Observer programs: G011180 (PI: Dressing) and G03278 (PI: Mayo).}.

In the following sections we describe first the initial candidate identification in the \tess data, followed by our own search for further candidates.


\subsection{\tess Candidate Identification}

All \tess 2-minute cadence data are processed in the first instance by the \textsc{SPOC} (Science Processing Operations Center) pipeline, presented in \cite{SPOC}. Data products are then available to the community in the form of Simple Aperture Photometry \citep[SAP][]{twicken:PA2010SPIE,morris:PA2020KDPH} or Presearch Data Conditioning Simple Aperture Photometry \citep[PDCSAP][]{Stumpe2012,Smith2012,Stumpe2014}, the latter having been corrected for instrument systematics and for crowding effects. Data products are downloadable from the NASA Mikulski Archive for Space Telescopes (MAST) and available via the \textsc{Lightkurve} package \citep{lightkurve}. All lightcurves are additionally searched by \textsc{SPOC} for periodic transit-like signals and candidates with $\rm SNR\,>\,7.1$ are reported as threshold crossing events (TCEs).

TOI-715.01 was first reported as a candidate on 2019\,May\,24 \citep{guerrero2021} following a multi-sector transit search \citep{jenkins2002,jenkins2010,jenkins2020} conducted on 2019\,May\,5 for sectors 1--9. The candidate was already identified in the multi-sector search of Sectors 1–-6 conducted on 2019\,April\,18, but did not at this time pass the necessary tests to be reported as a TOI. The transit signal was fitted with an initial limb-darkened transit model \citep{Li2019} and subjected to a suite of diagnostic tests \citep{Twicken2018} to help determine whether the signature is from an exoplanet. The transit signature passed all of the tests reported in the Data Validation report for Sectors 1-9\footnote{All data validation reports \citep{Twicken2018} referred to in this section are downloadable at \url{https://tev.mit.edu/data/search/?q=271971130}}, and the difference image centroiding test located the source of the transit signal to within $3\farcs7\pm3\farcs5$ in the subsequent data validation reports for the multi-sector searches of Sectors 1--13 and 27--39. 

At the time of the first reported TCE\footnote{Multi-sector TCE statistics available at \url{https://archive.stsci.edu/missions/tess/catalogs/tce/tess2018206190142-s0001-s0006_dvr-tcestats.csv}}, the \tess Input Catalog (TIC) in use was version 7, which had not yet incorporated the \textit{Gaia}-DR2 data; as such, TOI-715 did not have a stellar radius and the planet candidate was reported as a $\rm 7.1\,R_{\oplus}$ object with a period of $\rm \sim 19.2\,days$. Following the update to the TICv8 \citep{TICv8} the planet candidate's radius estimate was revised, demonstrating that TOI-715.01 was likely a super-Earth and thus a high priority target.

\textsc{SPOC} TCEs are subject to the \textsc{TESS-ExoClass} \footnote{\url{https://github.com/christopherburke/TESS-ExoClass}} automated classifier that reduces the number of TCEs that undergo the manual TOI vetting procedure. \textsc{TESS-ExoClass} applies a series of tests that are similar to the \textit{Kepler} \textsc{Robovetter} \citep{coughlin2016,thompson2018}. TOI 715.01 passes all the tests of \textsc{TESS-ExoClass} and has been placed in the Tier 1 (highest quality candidate) category, and has been classified as a Tier 1 candidate in the subsequent \textsc{SPOC} multi-sector 1--13, 1--36, and 1--39 searches. Further details for the TOI assignment process are available in \cite{guerrero2021}.

We present the PDCSAP \tess lightcurves for all 24 sectors of 2-minute cadence data in Figure \ref{fig:tess}; the timings of the 29 transits are indicated with dark pink arrows.

\subsection{Search for additional candidates}

We make use of the custom pipeline \textsc{Sherlock\footnote{\textsc{Sherlock} is publicly available at \url{https://github.com/franpoz/SHERLOCK}}}, presented in \citet{sherlockp1} and \citet{sherlockp2}, to search the \tess data for additional transiting candidates, as was done in \cite{TOI282}. \textsc{Sherlock} downloads all lightcurves from MAST and, using \textsc{Wotan} \citep{wotan}, applies a bi-weight function with varying window sizes to detrend the data. Each detrended lightcurve and the original PDCSAP lightcurve are then searched for transit-like signals using \textsc{Transit Least Squares} \citep{transitleastsquares}. We use only the 2-minute cadence in our candidate search, applying a Savitzky–Golay (SG) digital filter \citep{SVfilter} previously following the strategy described in \cite{speculoos2}\footnote{We use \textsc{SciPy}'s implementation of the SG filter, using a 3rd order polynomial and a window size of 11 points.} and we test 11 window sizes between 0.19 and 1.9 days when detrending with the bi-weight filter. We thus carry out our transit search on the PDCSAP lightcurve and the 11 detrended lightcurves and only consider signals with SNR>7 for further investigation.

We recover TOI-715.01 at a period of $\rm 19.29\,d$ in the first instance in all 12 lightcurves, and we find that the highest SNR and SDE are achieved in the PDCSAP flux without any detrending with \textsc{Wotan}.

\begin{figure}
    \centering
    \includegraphics[width=0.47\textwidth]{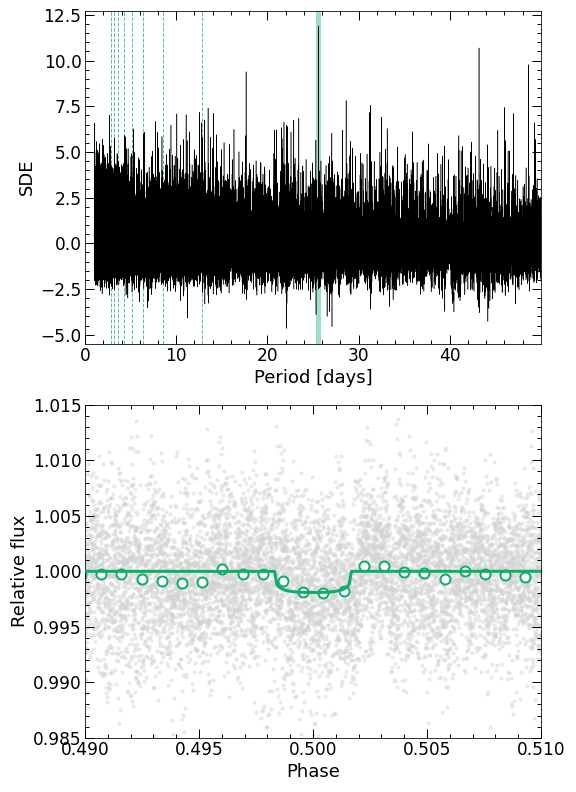}
    \caption{Results of our search for additional candidates in the data, using {\sc Transit Least Squares} as implemented by {\sc Sherlock}. \textbf{Upper panel:} \textsc{TLS} periodogram showing the detected 25.61 day period of TIC 271971130.02 and its harmonics. \textbf{Lower panel:} \tess PDCSAP lightcurve phase-folded on this period, with a transit model overplotted. We note that the errorbars on the binned points are smaller than the markers, and that the lightcurve shown here is the one treated with the SG Filter}.
    \label{fig:cand02}
\end{figure}

We additionally find two other periodic signals in the data. Of these, the signal with highest SNR is at a period of $\rm 25.61\,d$, putting it within $0.4\%$ of the first order 4:3 commensurability. The signal is detected in nine of the searched lightcurves with a maximum SNR of 13.81 and SDE of 11.6; it has a depth of $\rm \sim\,1\,ppt$, making this a $\rm\sim\,1.16\,R_{\oplus}$ candidate. In order to ensure this signal is not an artefact produced by the SG filter, we additionally search the untreated PDCSAP lightcurve and recover the signal with an SNR of 6.81. We adopt this candidate as TIC~271971130.02\footnote{This candidate has been submitted as a Community TESS Object of Interest (CTOI): \url{https://exofop.ipac.caltech.edu/tess/target.php?id=271971130}} and highlight the positions of transit events on Figure \ref{fig:tess} with dark green arrows. In Figure \ref{fig:cand02} we present the \tess lightcurve folded on this signal, along with the Lomb-Scargle periodogram. 

Given the large amount of data in Figure \ref{fig:tess}, it is not apparent by eye how much pre- and post-transit baseline each transit has. We therefore note that the total number of in-transit points for TIC~271971130.02 is 1371, while the total number of points before and after the transits are 1349 and 1369 respectively (counted up to 1 transit duration). Thus on average the pre-transit baseline is 98.4\% of a transit duration, while the post-transit baseline is 99.9\%.

The SPOC ran the Data Validation module at the ephemeris for the second signal and obtained an SNR of 5.5 sigma, and additionally identified other sub-threshold transit-like signals at higher SNR.

The second signal we recover has a period of $\rm 7.17\,d$ and a depth of $<1\,\rm ppt$; it has a maximum SNR of 9.63 and SDE of 8.17. The very low SNR of the signal makes it challenging to discern by eye whether or not the shape is consistent with that of a planetary transit. We also note that this signal is only found in four of the detrended lightcurves (with window sizes between 0.38--0.65 days, but not the PDCSAP lightcurve, indicating that the signal could be dependent on detrending. The duration of a transit at this period on a circular orbit is should be of order 0.065 days, and all tested windows are at least $3\times$ this duration. We therefore do not believe the detrending will have suppressed the transit in other lightcurves if it is real. When \tess returns to the southern skies in its second extended mission, the additional photometry will provide further insight into the nature of this signal.

It is important to note that the SNR and SDE yielded by \textsc{SHERLOCK} are inherited from the \textsc{Transit Least Squares} algorithm, which uses a simple estimation that can not be compared with values provided by other pipelines, such as SPOC. Instead, these values are used internally in \textsc{SHERLOCK} to compare the signals found and select the most prominent among them.

\section{Vetting and Validation}
\label{sec:followup}

In this section we describe the results of the multi-facility follow-up campaign conducted between May 2020 and April 2022. We begin with the high-resolution imaging observations, and then outline the photometric observations collected from five southern observatories. Finally, we describe how all our follow-up observations were used to validate the planetary nature of TOI-715.01 and TIC 271971130.02.

All follow-up observations are summarised in Table \ref{tab:followup}.

\begin{table*}
\centering
\caption{Summary of ground-based follow-up observations carried out for the validation of TOI-715.01 and TIC 271971130.02}.
\begin{tabular}{@{}ccccc@{}}
\midrule \midrule
\multicolumn{5}{c}{\textbf{Follow-up Observations}}                                              \\ \midrule \midrule
\multicolumn{5}{c}{\textbf{High Resolution Imaging}}                                             \\ 
\textbf{Observatory} & \textbf{Filter} & \textbf{Date}     & \textbf{Sensitivity Limit} & \textbf{Result}\\ \midrule
Gemini South      & $562~{\rm nm}$     & 2020 December 26 & $\Delta m=4.69$ at $0.5\arcsec$  & No sources detected  \\ 
Gemini South      & $832~{\rm nm}$     & 2020 December 26 & $\Delta m=5.07$ at $0.5\arcsec$  & No sources detected   \\ \midrule
\multicolumn{5}{c}{\textbf{Photometric Follow-up}}                                               \\
\textbf{Observatory} & \textbf{Filter} & \textbf{Date}     & \textbf{(Candidate) Coverage} & \textbf{Result} \\ \midrule
LCO-SAAO      & {\it Sloan-$i'$}     & 2020 May 13 & Ingress  & Transit ruled out on or off target during the time covered\\ 
LCO-SAAO      & {\it Sloan-$i'$}     & 2020 October 15 & (.01) Ingress  & Transit detected on target\\ 
ExTrA      & {\it $\rm 1.21\mu m$}     & 2021 February 26 & (.01) Full  & Detection\\
ExTrA      & {\it $\rm 1.21\mu m$}     & 2021 April 25 & (.01) Full  & Detection\\ 
TRAPPIST-South    & {\it $I+z'$}     & 2021 April 25 & (.01) Full  & Detection\\ 
LCO-CTIO    & {\it Sloan-$i'$}     & 2021 April 25 & (.01) Full  & Detection\\ 
SSO-Callisto    & {\it Sloan-$r'$}     & 2021 September 26 & (.01) Full  & Detection\\
SSO-Io    & {\it Sloan-$r'$}     & 2021 September 26 & (.01) Full  & Detection\\
SSO-Europa    & {\it Sloan-$r'$}     & 2021 September 26 & (.01) Full  & Detection\\
SSO-Ganymede    & {\it Sloan-$r'$}     & 2021 September 26 & (.01) Gapped  & Interruption due to weather - ingress detected\\
TRAPPIST-South    & {\it $I+z'$}     & 2021 September 26 & (.01) Full  & Detection\\
OACC-CAO    & \textit{Sloan-i'2}     & 2021 November 24 & (.01) Full  & Detection\\ 
SSO-Callisto    & {\it $I+z'$}     & 2021 November 24 & (.01) Full  & Detection\\
SSO-Io    & {\it $I+z'$}     & 2021 November 24 & (.01) Full  & Detection\\
SSO-Ganymede    & {\it $I+z'$}     & 2021 November 24 & (.01) Full  & Detection\\ 
TRAPPIST-South    & {\it $I+z'$}     & 2021 November 24 & (.01) Full  & Detection\\
ExTrA    & {\it $\rm 1.21\mu m$}     & 2022 February 08 & (.01) Full  & Detection\\
TRAPPIST-South    & {\it $I+z'$}     & 2022 April 07 & (.01) Full  & Detection\\
ExTrA    & {\it $\rm 1.21\mu m$}     & 2022 April 07 & (.01) Full  & Detection\\
TRAPPIST-South    & {\it $I+z'$}     & 2022 October 25 & (.02) Egress  & Inconclusive (high airmass)\\\midrule
\multicolumn{5}{c}{\textbf{Spectroscopic Observations}}                                               \\ 
\textbf{Instrument} & \textbf{Wavelength Range} & \textbf{Date}     & \textbf{Number of Spectra} & \textbf{Use}\\ \midrule
Magellan/LDSS3     & $380-1000~\rm nm$      &  2022 January 06 & 1  & Stellar characterisation  \\ \midrule

\end{tabular}
\label{tab:followup}
\end{table*}

\subsection{High resolution imaging}
\label{sec:hires}

Close stellar companions (bound or in the line of sight) can confound derived exoplanet  properties in a number of ways.  The detected transit signal might be a false positive due to a background eclipsing binary and even real planet discoveries will yield incorrect stellar and exoplanet parameters if a close companion exists and is unaccounted for \citep[e.g.,][]{Ciardi2015,FH2017,FH2020}. Additionally, the presence of a close companion star leads to the non-detection of small planets residing within the same exoplanetary system \citep{Lester2021}. 
Approximately 25\% of M dwarfs are part of binary or multiple star systems \citep[e.g.,][]{Mdwarf_mult_CARM2015,Mdwarf_mult_Winters2019}, though fewer than 5\% of known spectroscopic binaries include an M-dwarf primary star \citep{sb9}\footnote{\url{https://sb9.astro.ulb.ac.be}}. Nonetheless, high resolution imaging provides crucial information toward our understanding of exoplanetary formation, dynamics and evolution \citep{Howell2021}. 

\begin{figure}
    \centering
    \includegraphics[width =\columnwidth]{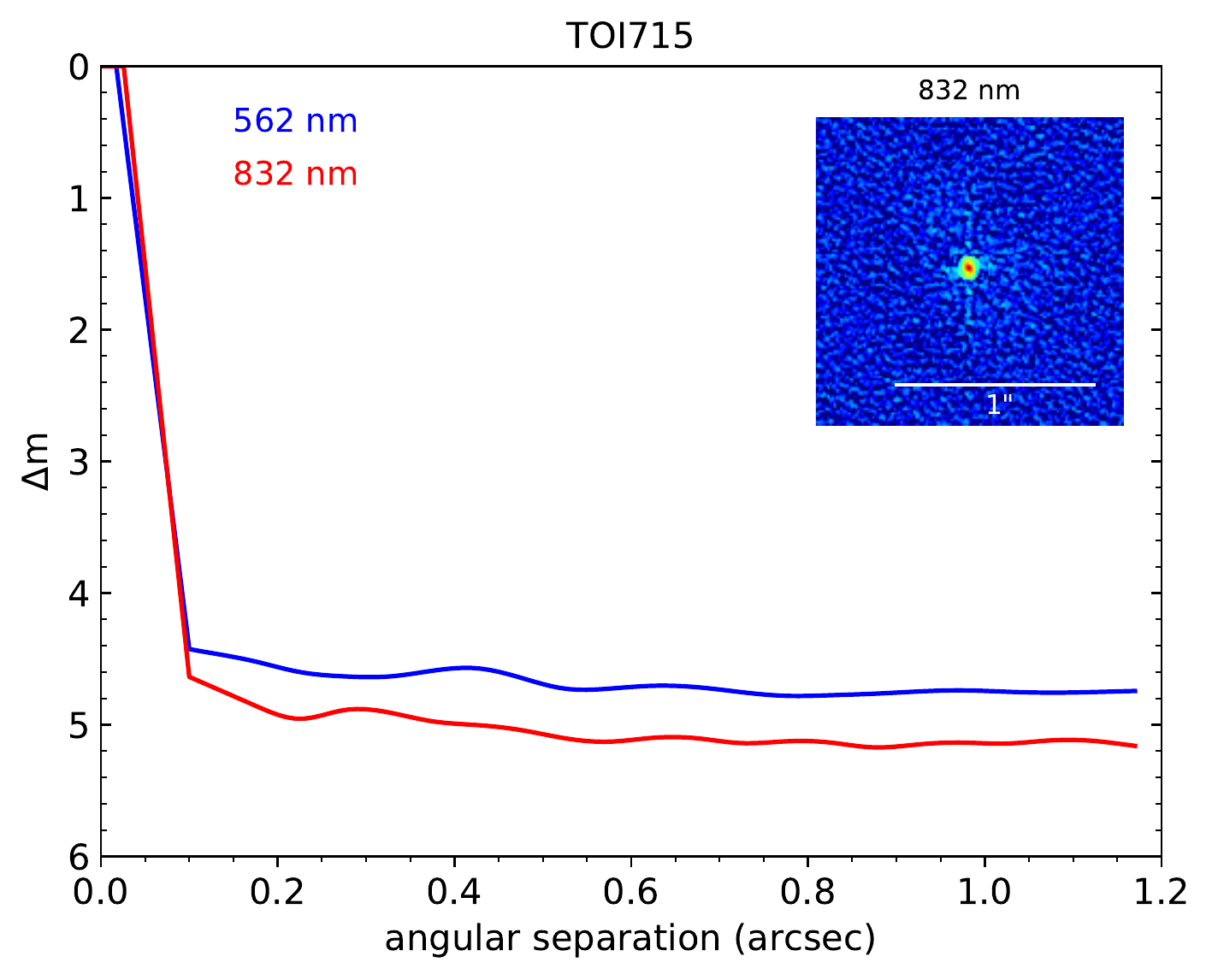}
    \caption{Plot showing the $5\sigma$ speckle imaging contrast curves in both filters as a function of the angular separation out to $1.2\arcsec$, the end of speckle coherence. The inset shows the reconstructed $832~\rm nm$ image with a $1\arcsec$ scale bar. The star, TOI-715, was found to have no close companions to within the angular and brightness contrast levels achieved.}
    \label{fig:hires}
\end{figure}

TOI-715 was observed on 2020\,December\,26 UT using the Zorro speckle instrument on the Gemini South 8-m telescope \citep{Scott2021}.  Zorro provides simultaneous speckle imaging in two bands ($562~\rm nm$ and $832~\rm nm$) with output data products including a reconstructed image with robust contrast limits on companion detection \citep[see][]{HF2022}. TOI-715 was found to be a single star to within the angular and brightness contrast levels achieved. Eight sets of $1000 \times 0.06$\,s images were obtained and processed by our standard reduction pipeline \citep{Howell2011}. Figure \ref{fig:hires} shows our final contrast curves and the $832~\rm nm$ reconstructed speckle image. These high-resolution observations revealed no companion star brighter than 5 magnitudes below that of the target star from the 8-m telescope diffraction limit ($20~\rm mas$) out to $1.2\arcsec$. At the distance of TOI-715 ($42.4~\rm pc$) these angular limits correspond to spatial limits of $0.84$ to $50.9~\rm AU$.

\subsection{Photometric follow-up}
\label{sec:photometry}

\subsubsection{Las Cumbres Observatory}

We used the Las Cumbres Observatory Global Telescopes \citep[LCOGT;][]{Brown:2013} 1.0\,m network nodes at South Africa Astronomical Observatory (SAAO) and Cerro Tololo Inter-American Observatory to observe three transits of TOI-715.01. First and second transits were observed with LCO-SAAO on 2020\,May\,13 and 2020\,October\,15, and third was observed with LCO-CTIO on 2021\,April\,25. All observations were carried out with the Sloan-$i'$ and an exposure time of 180\,s.

Photometric brightness was measured using an uncontaminated target aperture of 4.3\arcsec (11 pixels).
We used the {\sc TESS Transit Finder}, which is a customized version of the {\sc Tapir} software package \citep{Jensen:2013}, to schedule our photometric time series. The 1.0\,m telescopes are equipped with $4096\times4096$ SINISTRO cameras with an image scale of $0\farcs389$ per pixel, resulting in a $26\arcmin\times26\arcmin$ field-of-view. The raw images were calibrated with the standard LCO {\sc BANZAI} pipeline \citep{McCully:2018}, and photometric data were extracted with {\sc AstroImageJ} \citep{Collins:2017}.

The observation of 2020\,May\,13 only provided $\sim 20 \%$ in-transit coverage as well as 1.2 hours of pre-transit baseline. The transit was ruled out on or off target during the window covered by this observation\footnote{With the updated ephemeris we confirm that transit would have started just after the end of the window covered by this observation.}. The subsequent observation of 2020\,October\,15 confirmed the transit event on target with the detection of an ingress that was 29 minutes late relative to the ephemeris. The observation of 2021\,April\,25 resulted in the detection of a ful transit.

\subsubsection{SPECULOOS-Southern Observatory}

The SPECULOOS Southern Observatory (SSO) is comprised of four Ritchey-Chr\'etien 1.0\,m-class telescopes installed at ESO Paranal in the Atacama desert \citep{SSOscopes}. Designed to hunt for small, habitable-zone planets orbiting ultra-cool stars \citep{speculoos}, all four telescopes are equipped with a deep-depletion Andor CCD camera with $2048 \times 2048$ 13-$\micron$ pixels. Each telescope therefore has a field of view of $12\arcmin\,\times\,12\arcmin$ and a pixel scale of $0\farcs35$ \citep{SSOBook}.

All SPECULOOS observations are scheduled using the \textsc{python} package \textsc{SPOCK}\footnote{\url{https://github.com/educrot/SPOCK}} \citep{speculoos}, and processed in the first instance by an automatic data reduction pipeline, described in detail in \cite{SSOpipeline}. Successful observations of \textit{TESS} targets are then reprocessed using \textsc{prose}, a publicly available \textsc{python} framework for processing astronomical images\footnote{\url{https://github.com/lgrcia/prose}} as described in \citet{prosesoft,prosepaper}. Images are calibrated and aligned before performing aperture photometry on the 500 brightest sources detected; \textsc{prose} then performs differential photometry \citep{Broeg05} on the target star to extract the lightcurve. Individual lightcurves are detrended for airmass, sky background and FWHM (full width at half maximum) using second order polynomials in time, and they are then modelled using \textsc{exoplanet} \citep{exoplanet:joss}.

We observed TOI-715.01 for the first time on the night of 2021\,September\,26 with all four telescopes simultaneously (Io, Europa, Ganymede, Callisto) using the \textit{Sloan-$r'$} filter and $\rm 120$-s exposures. Cloudy skies during the transit caused some large systematics in three out of four telescopes, although this did not prevent the transit from being detected. The fourth telescope, Ganymede, closed during the weather alert, causing only the ingress to be detected. 

We re-observed TOI-715.01 on the night of 2021\,November\,24 with three of our telescopes. We collected $\rm 13$-s exposures using the custom filter $I+z'$ \citep{speculoos}. Observations were once again hampered by the untimely appearance of clouds, which affected the in-transit precision of our observations. Fortunately, the transit was still detected with all three instruments despite the overcast conditions.

\subsubsection{TRAPPIST-South}

We observed four transits of TOI-715.01 with TRAPPIST-South (TS) \citep{Jehin2011,Gillon2011}, located in ESO La Silla Observatory in Chile. This 0.6-m telescope is equipped with a FLI ProLine PL3041-BB camera and a back-illuminated CCD with a pixel size of $0\farcs64$, providing a total field of view of $22\arcmin\,\times\,22\arcmin$ for an array of $2048\,\times\,2048$ pixels. TS is a Ritchey-Chr\'etien telescope with F/8 and is equipped with a German equatorial mount.

All four transits were observed with the custom $I+z'$ filter to maximize the photometric precision. The observations took place on 2021\,April\,25, 2021\,September\,26, 2021\,November\,24 and 2022\,April\,07, with exposure times of 50s, 70s, 120s and 90s respectively. We reduced the images using \textsc{prose} pipeline \citep{prosepaper,prosesoft} to extract optimal light curves. Individual lightcurves were then modeled using the \textsc{exoplanet} \citep{exoplanet:joss} package and detrended using a second order polynomial of airmass, FWHM (full width at half maximum) and sky background. None of the observations suffered weather losses.

We additionally observed TOI-715 on 2022\,October\,25 during a partial transit of the second candidate (.02) but the target was too low during the event. The observation was therefore inconclusive.

\subsubsection{ExTrA}

ExTrA \citep{Bonfils2015} is a near-infrared (0.85--1.55~$\rm \mu m$) multi-object spectrograph fed by three 0.6\,m telescopes located at La Silla observatory. We observed four full transits of TOI 715.01 on 2021\,February\,26, 2021\,April\,25, 2022\,February\,08, and 2022\,April\,07. We observed the first three transits using two telescopes, and the fourth using three; all observations were carried out with $8\arcsec$ aperture fibers. For both nights, we used the spectrograph's low resolution mode ($R\sim20$) and 60s exposures. Five fibre positioners are used at the focal plane of each telescope to collect light from the target and four comparison stars. As comparison stars, we also observed 2MASS J07381369-7347351, 2MASS J07410628-7341297, 2MASS J07393394-7330535, and 2MASS J07372328-7321411 with J-magnitude \citep{Skrutskie2006} and $T_{\rm eff}$ \citep{gaia_edr3} similar to TOI-715. The resulting ExTrA data were analyzed with custom data reduction software, detailed in \cite{Cointepas2021}.

\subsubsection{OACC-CAO}

We observed a transit of TOI-715.01 with Campo Catino Austral Observatory (OACC-CAO), located in El Sauce Observatory in the Atacama desert in Chile, on the night of 2021\,November\,24. The telescope is a $\rm 0.6\,m$ Planewave CDK $24\arcsec$ with a Planewave L600 mount, equipped with an FLI filter wheel with \textit{Sloan} filters and an FLI PL16803 camera with $4096\times4096$ $9~\rm \mu m$ pixels. It has a field of view of $32\arcmin$ and a pixel scale of $0\farcs48$.

TOI-715.01 was observed with the \textit{Sloan-i'2} filter using 180s exposures. The images were reduced using \textsc{AstroImageJ} \citep{Collins:2017} and we detect the full transit in an uncontaminated aperture.

\subsection{Statistical Validation}
\label{sec:validation}


We make use of the statistical validation package \textsc{Triceratops}\footnote{Version 1.0.17.} \citep{triceratops,triceratops_soft} to assess the probability of the planet hypothesis for TOI-715.01 and TIC~271971130.02. \textsc{Triceratops} calculates the flux contribution from nearby stars to check if any could be responsible for the transit signal. It then calculates the relative probabilities of a range of transiting planet (TP) and eclipsing binary (EB) scenarios using lightcurve models fitted to the phase-folded photometry. 

In the first instance we find that the false positive probabilities (FPP) are 0.403 and 0.592 for candidates .01 and .02 respectively when examining the \tess data, while the threshold for validation is 0.015. For candidate .01, we make use of our follow-up photometry by folding in the four TRAPPIST-South lightcurves. We choose these as they were observed with the custom \textit{I+z'} filter (which is very similar to the \tess bandpass) and all four are at different epochs.

With this phase-folded subset of our follow-up photometry, we find that the false positive probability reduces to 0.0107, placing it below the threshold for statistical validation. To similarly reduce the FPP for TIC~271971130.02, we will need to collect ground-based photometry to confirm the event on-target or rule out events on nearby stars (such an observation was scheduled from Hazelwood Observatory but cancelled due to bad weather).

\subsubsection{Statistical validation conclusions}

 With a FPP of 0.0107 TOI-715.01 is now below the customary threshold for statistical validation \citep[0.015; ][]{triceratops}; we therefore consider the planet validated and refer to it as TOI-715\,b. TIC~271971130.02 does not yet meet the criterion for statistical validation, and is therefore classified as a `likely transiting planet' until more evidence such as ground-based photometry is collected.

\section{Global photometric analysis}
\label{sec:analysis}

\begin{figure*}
    \centering
    \includegraphics[width=0.9\textwidth]{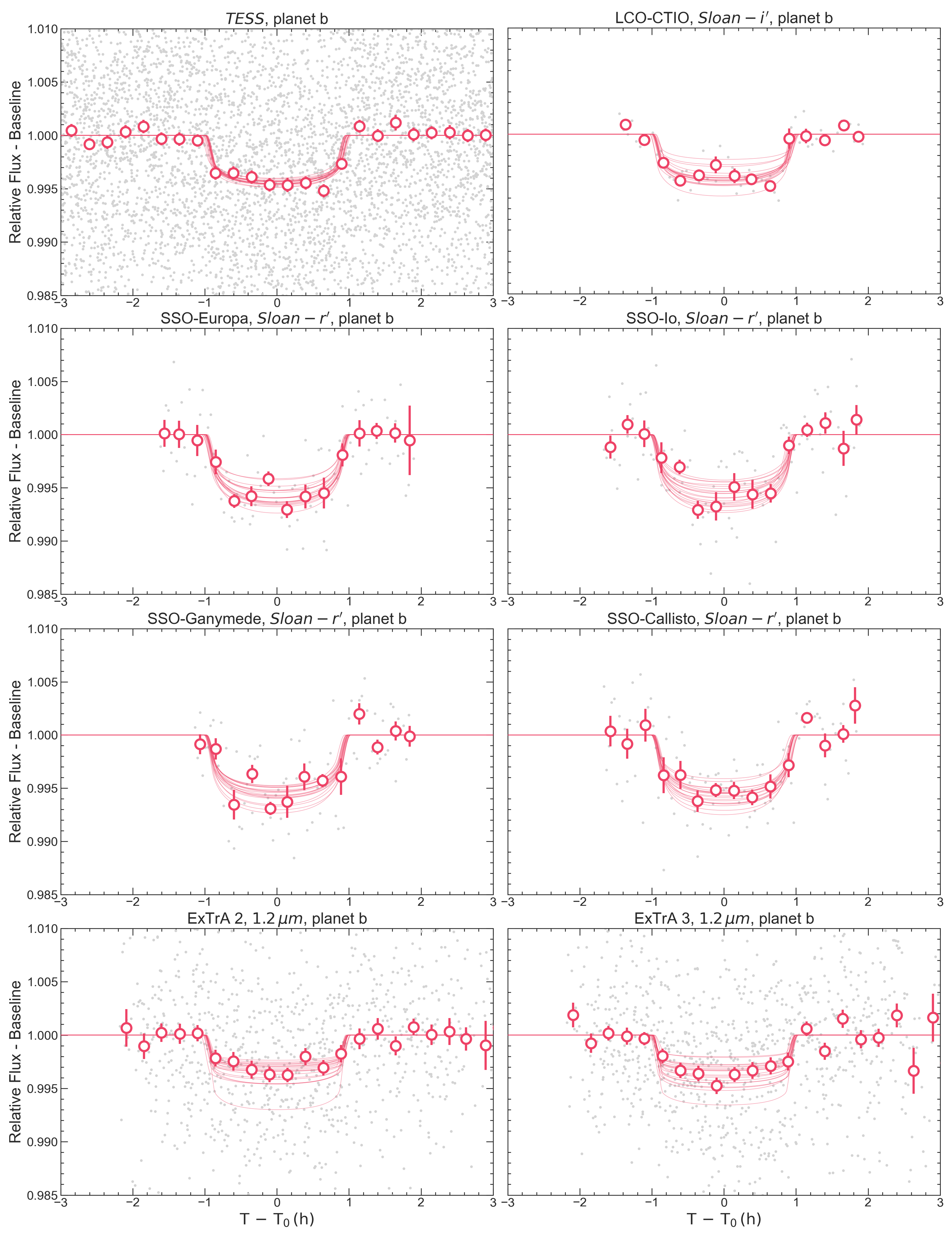}\
    \caption{Photometry of TOI-715\,b along with best fitting models. Grey points are raw flux and dark pink circles are 15-minute binned points. The dark pink lines are 20 fair draws from the posterior transit model. The flux and the transit models have been corrected by subtracting the baseline models. The \tess, ExTrA 2 and ExTrA 3 photometry are phase-folded, while the others are single transits.}
    \label{fig:transits_1}
\end{figure*}

\begin{figure*}
    \centering
    \includegraphics[width=0.9\textwidth]{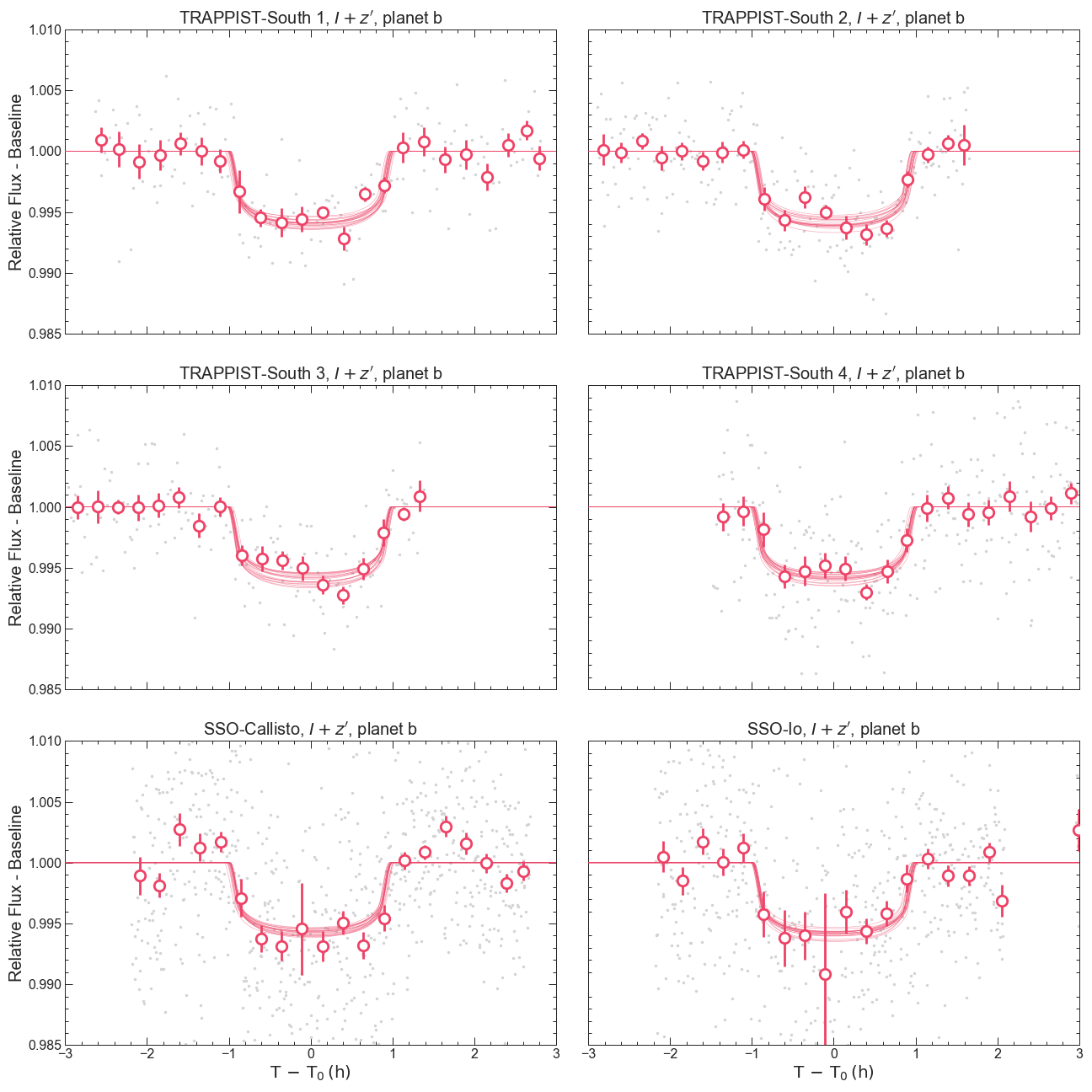}\
    \caption{Photometry of TOI-715\,b along with best fitting models. Grey points are raw flux and dark pink circles are 15-minute binned points. The dark pink lines are 20 fair draws from the posterior transit model. The flux and the transit models have been corrected by subtracting the baseline models. All transits shown here are single transits. For the TRAPPIST lightcurves, the number in the figure titles denotes the visit number. These are treated as single transits due to the different exposure times of each visit. Increased scatter during the transit observed by SSO-Io (bottom right panel) cause the middle binned point to have a large errorbar.}
    \label{fig:transits_2}
\end{figure*}

We carried out a global photometric analysis of the \tess photometry and the datasets described in Section \ref{sec:validation} using \textsc{Allesfitter} \citep{AllesfitterPaper,AllesfitterSoft}, a flexible and publicly available \textsc{python}-based inference package. We exclude all partial transits, as well as the transit obtained from OACC (due to large scatter in the data) from our analysis, leaving 15 full transits from our ground-based facilities. \textsc{Allesfitter} generates lightcurve models using \textsc{Ellc} \citep{ellc}, and Gaussian Process (GP) models using \textsc{Celerite} \citep{celerite}. The best fitting models are then chosen using either a nested sampling algorithm via \textsc{Dynesty} \citep{dynesty}, or MCMC sampling with \textsc{Emcee} \citep{emcee}. For all modeling in this paper, we make use of the nested sampling algorithm as it calculates the Bayesian evidence at each step in the sampling; this allows us to compare the Bayesian evidence for different models by calculating the Bayes Factor \citep{BayesFactor}. All prior distributions and their bounds can be found in Table \ref{tab:glob_fit}.

We adopt the signal parameters from Section \ref{sec:sherlock} as uniform priors, and the stellar parameters from Section \ref{sec:star} as normal priors and fit for all planetary parameters ($R_{\rm p}/R_\star$, $(R_\star+R_{\rm p})/a$, $\cos\,i$, $T_0$, and $P$). We fit two models in the first instance: one where eccentricity is constrained to zero and another where eccentricity is allowed to vary, parametrised as $\sqrt{e_{\rm b}} \cos{\omega_{\rm b}}$ and $\sqrt{e_{\rm b}} \sin{\omega_{\rm b}}$ {\citep[as in][]{Triaud2011}}.

The observations described in Section \ref{sec:validation} span several photometric filters from the blue to the near-infrared ends of the spectrum. To allow the colour-dependent transit depth to be a free parameter in our models, we also fit each lightcurve for a dilution parameter ($D$) which we give a uniform prior between -1 and 1. We fix the dilution of the \tess observations at 0 as the PDCSAP lightcurve is already corrected for crowding \citep{Stumpe2012}, and use the transit depth as the reference depth; we then use \textsc{Allesfitter}'s `$\rm coupled\_with$' functionality to ensure that the dilution and quadratic limb-darkening coefficients are fitted together for all observations taken in the same photometric band.

We use the \textsc{python} package \textsc{PyLDTK} \citep{pyldtk} and the PHOENIX stellar atmosphere library \citep{phoenix} to calculate quadratic limb-darkening coefficients for each photometric band. These are also adopted as normal priors in our models after we reparameterise them in the \cite{kippingldcs} parametrisation by converting from $u_1,u_2$ to $q_1,q_2$.

Finally, due to the complex instrument systematics present in the data obtained with the ExTrA telescopes, we model the correlated `red' noise in these observations using a GP. We select the \textit{Mat\'ern 3/2} kernel as implemented by \textsc{Celerite} due to its versatility and its ability to model both long and short term trends. We fit for two GP hyperparameters for each ExTrA telescope: the amplitude scale, $\sigma$, and the length scale, $\rho$. We place wide uniform priors on these terms. For all other telescopes, we make use of a hybrid spline to model the baseline.  \textsc{Allesfitter} also fits an `error scaling' term to account for white noise in the data.

\begin{figure}
    \centering
    \includegraphics[width=\columnwidth]{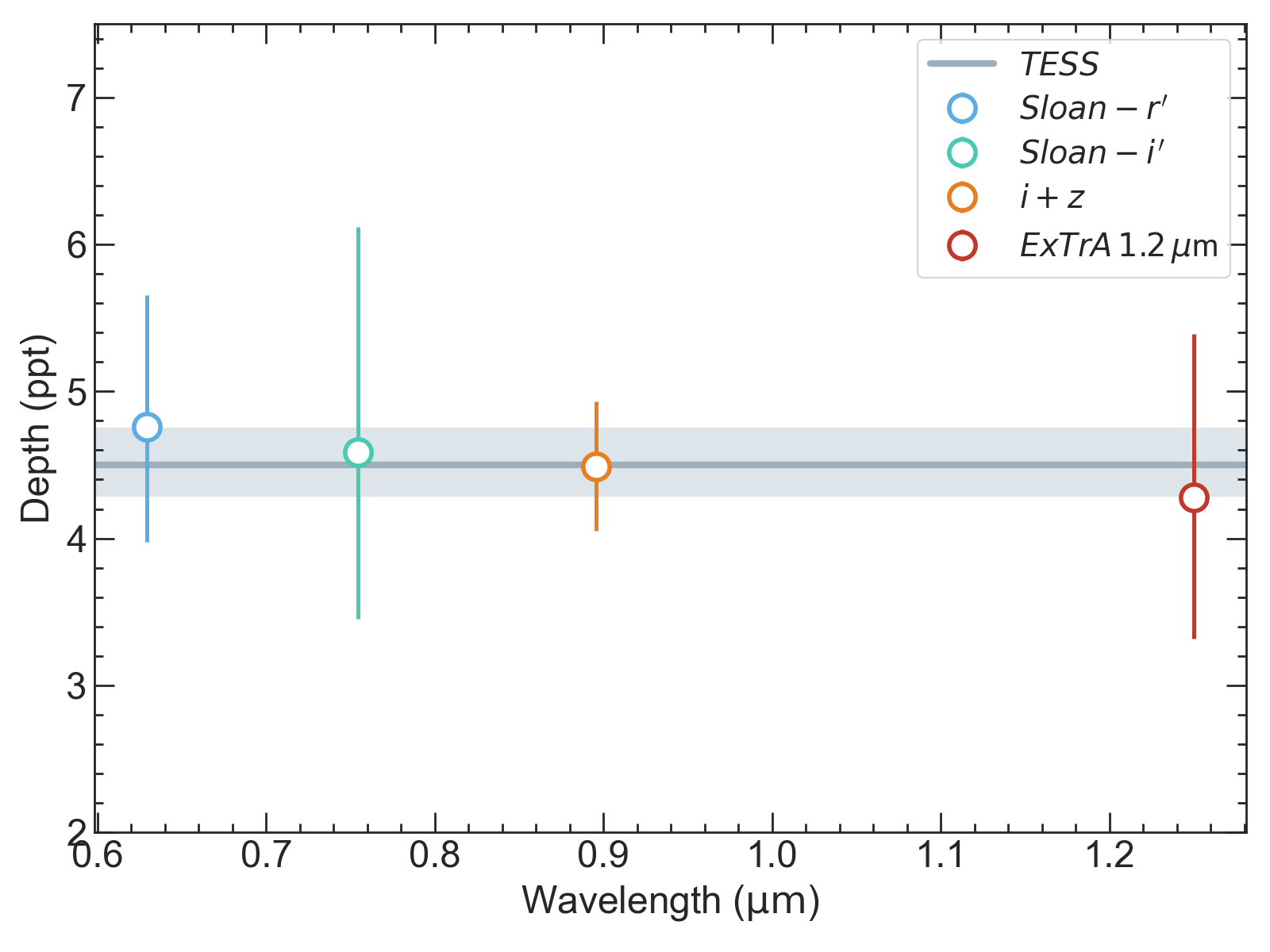}
    \caption{Measured transit depths vs. wavelength. The dark grey horizontal line indicates the depth of the \tess transits, while other depths are highlighted with circles for comparison.}
    \label{fig:depths}
\end{figure}

The initial fits (1-planet circular, and 1-planet eccentric) both confirm that TOI-715\,b is a $\rm 1.550\pm0.061\,\rm R_{\oplus}$ planet with an equilibrium temperature of $\rm 234\pm12\,\rm K$. We find that the eccentric model has slightly higher evidence, with a logged Bayes Factor of $\sim 6.5\pm1.1$, although the eccentricity derived from this more complex model,  $e = 0.31_{-0.20}^{+0.38}$, is consistent with zero at the $2\sigma$ level, and we do not consider this a detection of orbital eccentricity. All modelled ground-based transits of TOI-715\,b, as well as the phase-folded \tess photometry, can be found in Figures \ref{fig:transits_1} and \ref{fig:transits_2}. In Figure \ref{fig:depths} we present the averaged depths, in each photometric band, demonstrating the achromaticity of the observed transits.

We remind the reader that in Section \ref{sec:sherlock}, we identified a second transiting planet candidate in the \tess data, which we now incorporate into the model, and test this two-planet hypothesis. We test two further models to seek evidence of a second planet in the system, starting by fixing the linear ephemerides of planet b and allowing the mid-time of each individual transit to vary. This model is motivated by the second candidate's proximity to a $1^{\rm st}$ order 4:3 mean-motion resonance with the first planet. If the second candidate is real, we might expect for these planets to experience mutual gravitational interactions exciting transit timing variations (TTVs).

We place a wide uniform prior on each transit mid-time, corresponding to the linear predicted mid-time $\pm\, 60$ minutes.
We find that due to the low SNR of individual transits in the \tess data, the transit mid-times are not well constrained for events observed only by \tess. However, transits observed simultaneously by several ground-based observatories have significantly less scatter and smaller error bars. In these events, we see evidence suggesting TTVs of up to 5 minutes, although the Bayesian evidence for this model is lower than for the two linear models tested. Models with linear ephemerides are preferred, with Bayes Factors of $\sim9.5$ and $\sim16$ for the circular and eccentric models respectively. We note that we did not test a fixed linear ephemerides fit with a 2-planet model as the individual transits in \tess for the second candidate have even lower SNR and we have no ground-based transits at the time of writing.

\begin{figure}
    \centering
    \includegraphics[width=\columnwidth]{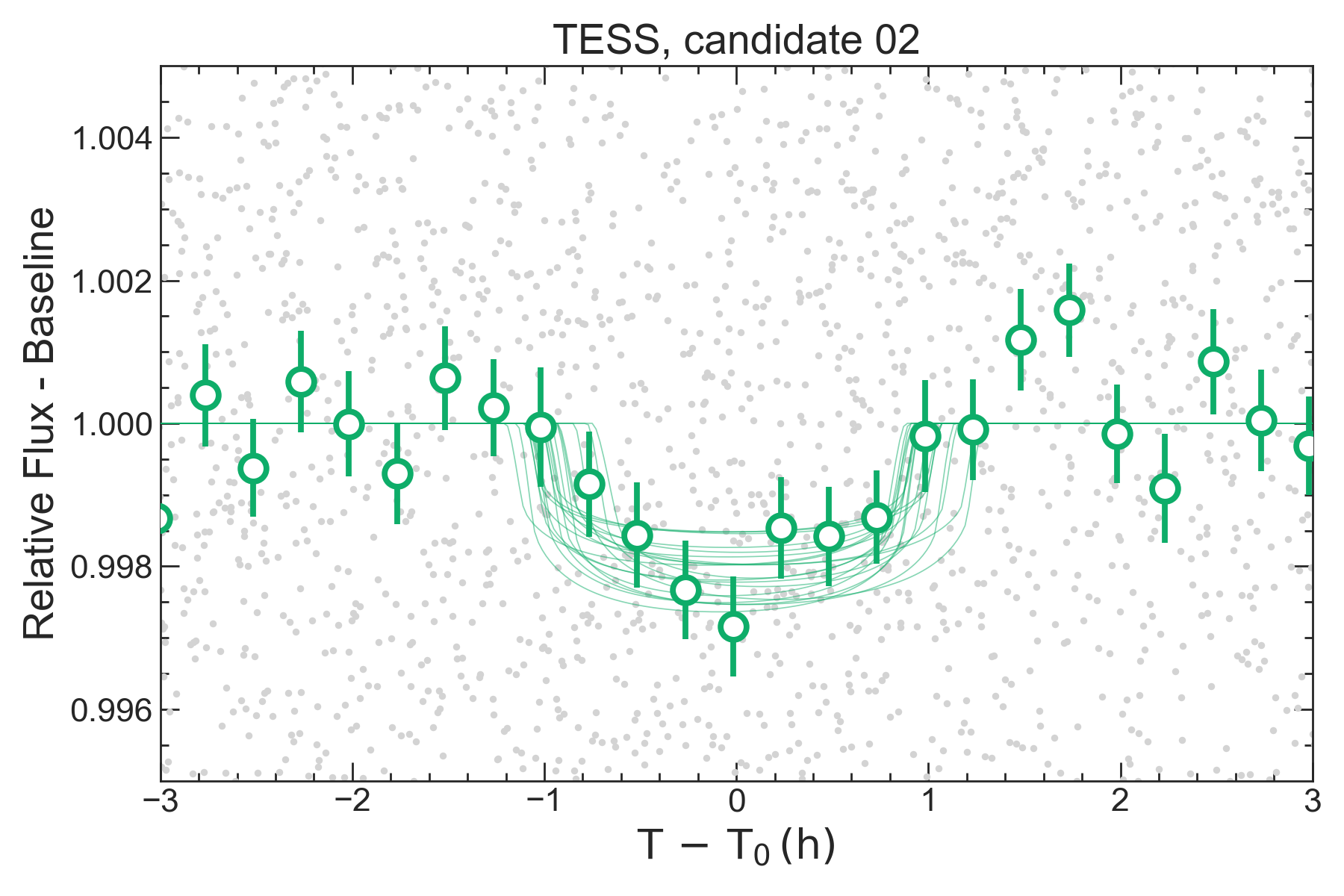}
    \caption{Phase folded \tess photometry of TIC~271971130.02. Grey points are raw flux and green circles are 15-minute binned points. The green lines are 20 fair draws from the posterior transit model. The flux and the transit models have been corrected by subtracting the baseline models.}
    \label{fig:planet_c}
\end{figure}

The final model we test is a 2-planet circular Keplerian model. We fit for all the same parameters as before, but now additionally include the transit parameters for the second candidate. The phase-folded \tess photometry for the second candidate is presented in Figure \ref{fig:planet_c}; the fitted and derived parameters for planet b resulting from the 2-planet model are consistent with those emerging from all previous models. Also noteworthy is that the host density as calculated by \textsc{Allesfitter} following \cite{Seager2003} with the transit parameters of the second candidate is $\rm22.7\pm3.0\,\rm g\,cm^{-3}$, which is within $1\sigma$ of the stellar density prior of $\rm 23.1\pm 3.2\,\rm g\,cm^{-3}$, suggesting that both signals are produced over the same star. The results of this fit indicate that if TIC~271971130.02 is confirmed, it is likely a $1.066\pm0.092 ~\rm R_\oplus$ planet with an orbital period of $25.60712_{-0.00036}^{+0.00031}$ and an equilibrium temperature of $215\pm12~\rm K$.

All fitted and derived parameters from the linear 2-planet circular model are presented in Table \ref{tab:glob_fit}.

\begin{table*}
\centering
\caption{Priors used in our fit, along with fitted and derived parameters. Uniform priors are indicated as $\rm \mathcal{U}(lower~bound,~upper~bound)$ and normal priors are indicated as $\rm \mathcal{N}(mean,~standard~deviation)$. $^\star$Equillibrium temperature is calculated assuming an albedo of 0.3 and emissivity of 1.}
\resizebox{\textwidth}{!}{%
\begin{tabular}{@{}cccc@{}}
\toprule
{ \bf }                      & { \bf TOI-715~b}                       & { \bf TIC~271971130.02}        &      \\ \midrule \midrule
\multicolumn{4}{c}{\textbf{Fit Parametrisation and Priors}}                           \\ \midrule
Transit Depth; $ R_{\rm p} / R_\star$         & $\mathcal{U}(0.01, 0.1)$   & $\mathcal{U}(0.01, 0.1)$   \\
Inverse Scaled Semi-major Axis; $(R_\star + R_{\rm p}) / a$   & $\mathcal{U}(0.01, 0.05)$    & $\mathcal{U}(0.008, 0.04)$    \\
Orbital Inclination; $\cos{i}$               & $\mathcal{U}(0.000, 0.04)$    & $\mathcal{U}(0.000, 0.04)$   \\
Transit Epoch; $T_{0}$ (BJD)  & $\mathcal{U}(2\,458\,327.3, 2\,458\,327.7)$   & $\mathcal{U}(2\,458\,342.0, 2\,458\,342.4)$  \\
Period; $P$  (days)          & $\mathcal{U}(19.0, 19.4)$    & $\mathcal{U}(25.4, 25.8)$   & \\\midrule
\multicolumn{4}{c}{\textbf{Limb Darkening Coefficients}} \\ \midrule
\tess $\rm u_1$ & $0.3322\pm0.0013$ & \tess $\rm q_1$ & $\mathcal{N}(0.413, 0.050)$ \\
\tess $\rm u_2$ & $0.3103\pm0.0047$ & \tess $\rm q_2$ & $\mathcal{N}(0.259, 0.050)$  \\
\textit{I+z} $\rm u_1$ & $0.2942\pm0.0014$ & \textit{I+z} $\rm q_1$ & $\mathcal{N}(0.453, 0.050)$ \\
\textit{I+z} $\rm u_2$ & $0.3791\pm0.0052$ & \textit{I+z} $\rm q_2$ & $\mathcal{N}(0.218, 0.050)$  \\
\textit{Sloan-i'} $\rm u_1$ & $0.3863\pm0.0018$ & \textit{Sloan-i'} $\rm q_1$ & $\mathcal{N}(0.532, 0.050)$ \\
\textit{Sloan-i'} $\rm u_2$ & $0.3431\pm0.0060$ & \textit{Sloan-i'} $\rm q_2$ & $\mathcal{N}(0.265, 0.050)$  \\
\textit{Sloan-r'} $\rm u_1$ & $0.6408\pm0.0032$ & \textit{Sloan-r'} $\rm q_1$ & $\mathcal{N}(0.768, 0.050)$ \\
\textit{Sloan-r'} $\rm u_2$ & $0.2273\pm0.0079$ & \textit{Sloan-r'} $\rm q_2$ & $\mathcal{N}(0.218, 0.050)$  \\
\textit{ExTra (1.2$\mu$m)} $\rm u_1$ & $0.2058\pm0.0007$ & \textit{ExTra (1.2$\mu$m)} $\rm q_1$ & $\mathcal{N}(0.142, 0.050)$ \\
\textit{ExTra (1.2$\mu$m)} $\rm u_2$ & $0.1708\pm0.0028$ & \textit{ExTra (1.2$\mu$m)} $\rm q_2$ & $\mathcal{N}(0.273, 0.050)$  \\ \midrule
\multicolumn{2}{c}{\textbf{External Priors}}  & \multicolumn{2}{c}{\textbf{GP Priors}}                              \\\midrule
Stellar Mass; $M_{\star}$ ($\mathrm{M_{\odot}}$)  & $\mathcal{N}(0.225, 0.012)$ & Amplitude Scale $\mathrm{GP \ln \sigma (flux)}$ & $\mathcal{U}(-7, -7)$      \\
Stellar Radius;  $R_{\star}$ ($\mathrm{R_{\odot}}$)  & $\mathcal{N}(0.240, 0.012)$  & Length scale $\mathrm{GP \ln \rho (flux)}$   & $\mathcal{U}(-7, 2)$  \\
Stellar Effective Temperature; $T_{\rm eff}$ (K) & $\mathcal{N}(3075, 75)$ &                     &              \\\midrule
\multicolumn{3}{c}{{ \bf Fitted Parameters}}  & \textcolor{black}{\textbf{Source}}                      \\ \midrule
$R_{\rm p} / R_\star $      & $0.0618\pm0.0017$ &    $0.0425\pm0.0034$     & 2-planet linear fit     \\
$(R_\star + R_{\rm p}) / a$         & $0.01367\pm0.00017$ & $0.01129_{-0.00047}^{+0.00055}$   & 2-planet linear fit \\
$\cos{i}$       & $0.00252_{-0.00032}^{+0.00030}$ & $0.0048_{-0.0024}^{+0.0021}$       &  2-planet linear fit    \\
$T_{0}$ $(\mathrm{BJD})$    & $2\,459\,002.63051_{-0.00074}^{+0.00070}$ & $2\,459\,007.9879_{-0.0041}^{+0.0045}$   & 2-planet linear fit \\
$P$ $(\mathrm{d})$       & $19.288004_{-0.000024}^{+0.000027}$ &  $25.60712_{-0.00036}^{+0.00031}$       & 2-planet linear fit  \\ \midrule
\multicolumn{3}{c}{\bf Fitted GP Hyperparameters} &  \textcolor{black}{\textbf{Source}}  \\ \midrule
ExTrA 2  & $\mathrm{gp \ln \sigma}$  $-3.77_{-0.19}^{+0.24}$ & $\mathrm{gp \ln \rho }$ $-2.73_{-0.17}^{+0.26}$  & 2-planet linear fit \\
ExTrA 3  & $\mathrm{gp \ln \sigma}$  $-4.60_{-0.19}^{+0.22}$ & $\mathrm{gp \ln \rho }$ $-2.77_{-0.15}^{+0.24}$  & 2-planet linear fit \\ \midrule
\multicolumn{3}{c}{{ \bf Derived Parameters}} & \textcolor{black}{\textbf{Source}}        \\ \midrule
Companion radius; $R_\mathrm{p}$ ($\mathrm{R_{\oplus}}$)               & $1.550\pm0.064$       &   $1.066\pm0.092$                       & 2-planet linear fit  \\
Instellation; $S_\mathrm{p}$ ($\mathrm{S_{\oplus}}$)               & $0.67_{-0.20}^{+0.15}$       &   $0.48_{-0.17}^{+0.12}$                       & 2-planet linear fit  \\
Semi-major axis; $a$ (AU)                                              & $0.0830\pm0.0027$          & $0.0986\pm0.0054$                    & 2-planet linear fit \\
Inclination; $i$ (deg)                                                 & $89.856_{-0.017}^{+0.018}$     & $89.72_{-0.12}^{+0.14}$             & 2-planet linear fit \\
Impact parameter; $b$                                                  & $0.195_{-0.024}^{+0.023}$    & $0.44_{-0.22}^{+0.18}$          & 2-planet linear fit  \\
Total transit duration; $T_\mathrm{tot}$ (h)                           & $1.980\pm0.025$        & $1.99_{-0.21}^{+0.14}$                    & 2-planet linear fit  \\
Full-transit duration; $T_\mathrm{full}$ (h)                           & $1.741\pm0.022$        & $1.79_{-0.25}^{+0.15}$                    & 2-planet linear fit  \\
Equilibrium temperature$^{\star}$; $T_\mathrm{eq}$ (K)                           & $234\pm12$                   & $215\pm12$               & 2-planet linear fit  \\
Transit depth \tess; $\delta_\mathrm{tr; TESS}$ (ppt)   &    $4.48_{-0.23}^{+0.26}$   & $2.03_{-0.27}^{+0.31}$    & 2-planet linear fit      \\
Transit depth \textit{i+z}; $\delta_\mathrm{tr; i+z}$ (ppt) &  $4.48_{-0.48}^{+0.54}$     &     -           & 2-planet linear fit  \\
Transit depth \textit{Sloan-i'}; $\delta_\mathrm{tr; Sloan-i'}$ (ppt)   &    $4.60_{-1.20}^{+1.50}$   &          -    & 2-planet linear fit      \\
Transit depth \textit{Sloan-r'}; $\delta_\mathrm{tr; Sloan-r'}$ (ppt)   &    $4.81_{-0.90}^{+0.78}$   &  -   & 2-planet linear fit      \\
Transit depth \textit{ExTra} ($1.2~\rm \mu m$); $\delta_\mathrm{tr; ExTra (1.2\mu m)}$ (ppt)   &    $4.20_{-0.96}^{+1.10}$   &    -    & 2-planet linear fit      \\

\end{tabular}%
}
\label{tab:glob_fit}
\end{table*}

\section{Discussion}
\label{sec:discussion}

TOI-715 is host to at least one planet (TOI-715\,b), with $R_{\rm b}= \rm 1.550\pm0.064\,R_{\oplus}$, receiving an instellation $S_{\rm b} = \rm 0.67_{-0.20}^{+0.15}\,S_{\oplus}$, which places it within the `conservative habitable zone' defined by \citet{2013ApJ...765..131K} as the circumstellar region where a rocky planet recieves and instellation flux of $0.42-0.842\,S_{\oplus}$. 

Our model suggests that there is also a possibly second, smaller planet in the system, with size $R_{02} = 1.066\pm0.092 ~\rm R_\oplus$, just within the outer edge of the host's habitable zone, at an instellation $S_{\rm 02} = \rm 0.48_{-0.17}^{+0.12}\,S_{\oplus}$. 

Expectation of planet yield by \tess prior to launch showed \tess would detect of order $70\pm9$ Earth-sized planets ($<1.25 ~\rm R_\oplus$) and $14\pm4$ conservative `habitable zone' planets $<2~\rm R_\oplus$ \citep[as in][]{2013ApJ...765..131K}, but that \tess would identify of order $0$ Earth-sized ($<1.25 ~\rm R_\oplus$) `conservative habitable zone planets'  \citep{Sullivan2015}. More recent yield estimates (post-launch) by \cite{Kunimoto2022} appear more pessimistic, however, finding only $5\pm2$ `conservative habitable zone planets' $<2~\rm R_\oplus$ would be detected within the primary mission and first extended mission of {\it TESS}, a factor nearly three times lower than \citet{Sullivan2015}. Should the second planet candidate, TIC~271971130.02 be fully confirmed, its existence would thus represent an unexpectedly important discovery made possible by the \tess mission, with the contributions of many ground-based facilities.

In this section, we contextualise the system by examining other factors required for habitability, and assessing its suitability for further in-depth characterisation. We begin by revisiting the radius valley to understand what can be learned about the mass of TOI-715\,b from current mass-radius relations. We then explore the prospects for a precise mass measurement of this planet with current spectroscopic instrumentation, followed by a discussion of the potential for atmospheric characterisation with \textit{HST} and \textit{JWST}.

\subsection{TOI-715 b and the radius valley}
\label{per-rad}

The radius valley, as identified in the \textit{Kepler} sample by \cite{Fulton2017}, shows a lack of close-in ($P<100~\rm d$) planets with sizes between $1.5-2~\rm R_{\oplus}$; TOI-715\,b's $1.55~\rm R_\oplus$ appears just inside this underpopulated region of parameter space. The importance of the radius valley lies in its potential to teach us about planetary formation and post-formation evolution \citep[e.g.][]{Owen2017,Ginzburg2018,venturini202}, and hence planets inside this gap are crucial in furthering our understanding of the factors that sculpt it. 

The location and slope of the radius valley has been shown to depend on the properties of the host star \citep[e.g.][]{Fulton2018b,gupta2019,Berger2020}, and \cite{cloutier2020} and \cite{vaneylen2021} recently re-examined this for low-mass stars. \cite{cloutier2020} looked at the small, close-in planet distribution around stars cooler than $4700\,\rm K$ and found that the slope is opposite in sign compared with FGK stars, and also that the peaks of the planet radius distributions are shifted toward smaller planets. According to this, the center of the radius valley also shifts towards smaller radii. The sample examined in this study contained planets discovered by \textit{Kepler} and \textit{K2}, and was corrected for completeness. This work states that planets falling between their measured radius valley and the one measured by \cite{martinez2019} for Sun-like stars are so-called `keystone planets', crucial for further understanding the radius valley around low-mass stars. By this metric TOI-715\,b's radius of $\rm 1.55\,R_{\oplus}$ is just below the boundary of $\rm 1.66\,R_{\oplus}$ for a period of 19.288 days, falling within the super-Earth category.

\cite{vaneylen2021}, however, used a significantly smaller sample, restricted to confirmed and well-characterised planets, with masses and radii measured to at least 20\% precision, orbiting stars cooler than 4000\,K. They found a slope in radius-period space opposite in sign compared with \cite{cloutier2020}, and therefore consistent in sign with FGK stars as measured by, for instance, \cite{martinez2019}. They do, however, find that the valley shifts towards smaller radii for later spectral types. Contrary to the work of \cite{cloutier2020}, the functional form of the radius valley derived in this study places TOI-715\,b above the radius valley and with the population of sub-Neptunes.

\begin{figure}
    \centering
    \includegraphics[width=\columnwidth]{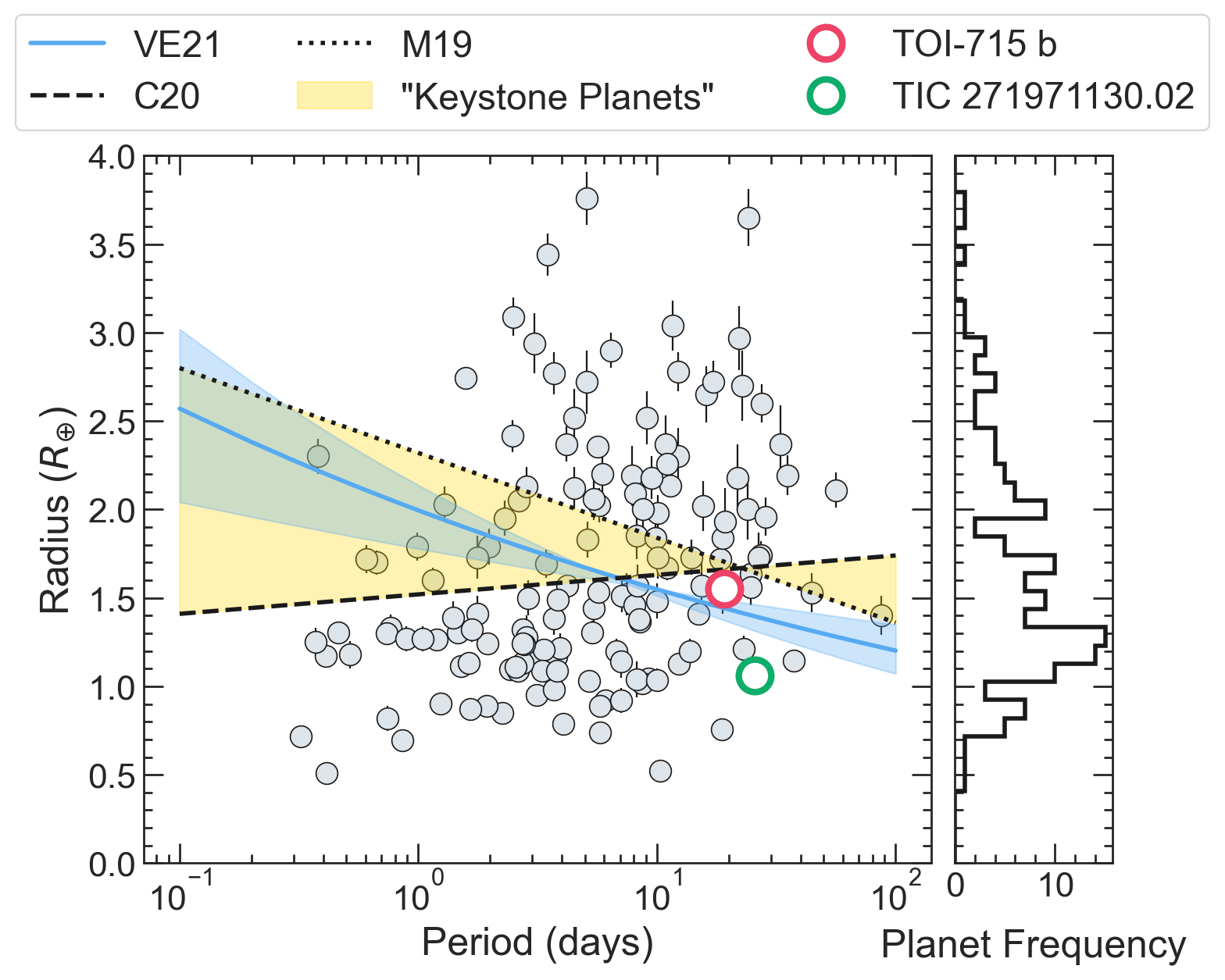}
    \caption{Plot of the current sample of planets orbiting hosts cooler than 4000\,K in radius-period space. The grey circles are confirmed planets with radii measured to better than 10\% precision, and the dark pink and green circles are TOI-715\,b and TIC 271971130.02 respectively. We note that some errorbars are smaller than the markers. The blue line indicates the measured radius valley according to \protect\cite{vaneylen2021} for hosts cooler than 4000\,K. The black dashed line is the low-mass star radius valley as measured by \protect\cite{cloutier2020} for hosts cooler than 4700\,K, while the dotted black line indicated the radius valley for Sun-like stars as measured by \protect\cite{martinez2019}. The yellow shaded area indicates where planets referred to as `keystone planets' can be found. The histogram in the right-hand panel does not include TOI-715\,b or TIC 271971130.02.}
    \label{fig:rad_val}
\end{figure}

In Figure \ref{fig:rad_val} we present the current sample of exoplanets orbiting stars cooler than $T_\mathrm{eff}\,=\,4000\, \rm K$ with planetary radii measured to better than 10\% precision\footnote{Retrieved from the NASA Exoplanet Archive on 2022 October 05, \url{https://exoplanetarchive.ipac.caltech.edu/cgi-bin/TblView/nph-tblView?app=ExoTbls&config=PSCompPars}} in radius-period space. We highlight the positions of the radius valleys as measured by \cite{martinez2019,cloutier2020} and \cite{vaneylen2021} and the positions of TOI-715\,b and TIC~271971130.02. In this sample, a clear radius bimodality for planets orbiting M dwarfs is not visible. What is evident is that precise characterisation of planets such as TOI-715\,b that fall \textit{between} the various definitions of the low-mass star radius valley is essential to understand whether or not this bimodality will eventually be borne out by the data.

\subsection{TOI-715 b and the density gap}

\cite{Luque2022} recently demonstrated using a sample of planets with precisely measured radii (to at least <8\%) and masses (to at least <25\%) that small planets orbiting low-mass stars in fact likely have a density gap rather than a radius valley. The results of that study indicate that small planets come in three flavours: rocky, water-worlds and gassy. For planets smaller than $2\,\rm R_{\oplus}$ they find a clear bi-modality in density which allow us to estimate the mass of TOI-715\,b in both the rocky ($\rm \sim\,7M_{\oplus}$) and water-world ($\rm \sim\,2M_{\oplus}$) scenarios; both different from the \cite{2017ApJ...834...17C} estimate of $\rm \sim\,3.5M_{\oplus}$ and the rocky and volatile-rich predictions from \cite{otegi2020} ($\rm \sim\,4.1M_{\oplus}$ and $\rm \sim\,3.5M_{\oplus}$ respectively). While this result is very promising for our understanding of the composition of small planets around low-mass stars, their sample contained only 34 planets; therefore continued systematic characterisation of these systems with consistent sample selection criteria remains crucial to understand this distribution and ultimately the composition of planets around M dwarfs.

\subsection{Prospects for a mass measurement of TOI-715 b}
\label{sec:mass}

Given the faintness of TOI-715, the only southern hemisphere instrument currently capable of the precision required to measure the mass of planet b is ESPRESSO at the VLT \citep[Very Large Telescope][]{pepe10}. If the planet is rocky and has a mass of $\rm \sim 7\,M_{\oplus}$, then we calculate a predicted RV semi-amplitude of $\rm 4.5\,m\,s^{-1}$. From ESPRESSO's exposure time calculator (ETC) we find that we could achieve a precision of $\rm 7.4\,m\,s^{-1}$ per measurement, meaning that for a 25\% precision on the mass of the planet, 44 spectra need to be collected, with $\rm 1800\,s$ exposures and SNR $\rm \sim$ 14. For this, we assume that TOI-715 is a relatively slow rotator ($\rm vsini \lesssim 2\,km\,s^{-1}$) which is not unusual for old M dwarfs that do not show significant stellar activity \citep{2017MNRAS.472.4563M,2022A&A...662A..41R}.

However the ETC is usually pessimistic. Based on observed radial-velocity precision of mid-M type planet-host stars obtained with ESPRESSO we expect a slightly better performance for this instrument. Observations of Proxima Cen (M5V) resulted in a mean precision of $\rm 0.6\,m\,s^{-1}$ \citep{2020A&A...639A..77S}, which is about 40\% better than the ETC predicts. A similar case is LHS 1140 (M3V), for which a mean precision of $\rm 0.8\,m\,s^{-1}$ has been achieved \citep{2020A&A...642A.121L}, which is a precision twice better than expected. Empirical results\footnote{\href{https://www.gemini.edu/instrumentation/maroon-x/exposure-time-estimation\#slit}{Empirical RV Uncertainties - MAROON-X}} for M dwarfs using the MAROON-X spectrograph at Gemini North \citep{2020SPIE11447E..1FS} found a more than 40\% increase in precision between M0 and M6 dwarf stars. The ESPRESSO ETC assumes an M2 spectral type to predict the RV precision. Thus, this apparent increase in RV precision for mid-M dwarfs, compared to early M dwarfs can explain the observed difference in performance.

Since TOI-715 is an M4 dwarf, we can conservatively expect a 20\% better RV precision compared to the ETC (thus a precision of $\rm \sim 5.9\,m\,s^{-1}$). This means a 25\% precision on the mass of the planet can be reached with only $\rm \sim 30$ spectra ($\rm 1800\,s$ exposures; SNR $\sim$ 14). This amounts to 15 hours of telescope time, which is not unrealistic for a planet of this importance.

If however, TOI-715\,b is in fact a `water world', then a mass of $\rm \sim2\,M_{\oplus}$ would produce an RV semi-amplitude of just $\rm 1.3\,m\,s^{-1}$. 
In this case 150 hours of telescope time are needed to constrain the mass with 25\% precision. Should the density bi-modality for small planets around M dwarfs suggested by \cite{Luque2022} be confirmed, a non-detection might enable us to infer the planet's mass.

In Section \ref{sec:analysis} we found that there were small hints of transit timing variations (TTVs) in the ground-based transits of TOI-715\,b. If the second planet is confirmed and the orbits are commensurate, the masses of both planets could be suitable for characterisation using dynamical modelling. Given the low SNR of individual transits and the long orbital periods, such a campaign would be well suited to ASTEP \citep[Antarctic Search for Transiting Exoplanets,][]{astep400,astep+}, given its convenient Antarctic location \citep{ASTEP}.

\subsection{Prospects for detailed atmospheric characterisation of TOI-715~b}
\label{sec:atmos}

\begin{figure}
    \centering
    \includegraphics[width=\columnwidth]{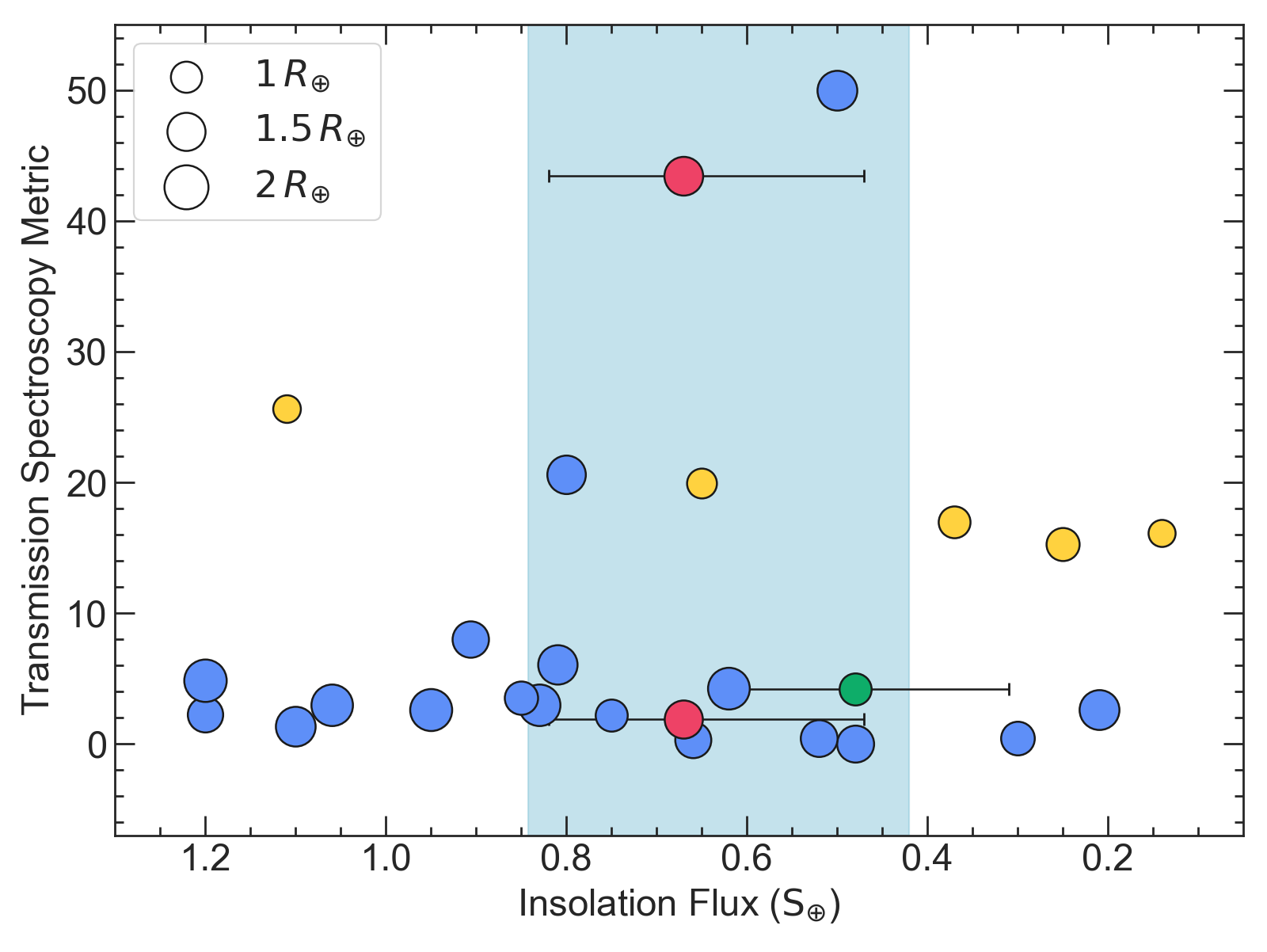}
    \caption{Transmission spectroscopy metric vs. instellation for transiting planets orbiting hosts cooler than 4000 K. The blue shaded area shows the conservative habitable zone as defined by \protect\cite{2013ApJ...765..131K}. The two dark pink circles highlight the positions of TOI-715 b in the rocky and water world scenarios, while the green circle indicates the position of TIC~271971130.02. We highlight in yellow the positions of the seven planets of TRAPPIST-1. The sizes of the circles are scaled with planetary radius.}
    \label{fig:TSM}
\end{figure}

The Transmission Spectroscopy Metric (TSM) as established by \cite{kempton2018} is often used to quantify the suitability of planets for atmospheric characterisation via transmission spectroscopy. We calculate the TSM for TOI-715 b in both mass limits\footnote{The formula for TSM calculations includes a planetary radius-dependent scale factor, with the cutoff for `terrestrial planets' at 1.5\,R$_{\oplus}$. As the radius of TOI-715\,b is $1.55\,\rm R_{\oplus}$ we use the terrestrial scale factor (0.190) for the `rocky world' TSM, and the `sub-Neptune' scale factor (1.26) for the `water world' TSM to enable a fair comparison.} and for TIC~271971130.02, as well as the published sample of planets with $R_p\,<\,2~\rm R_{\oplus}$ and $S_p\,<\,1.6~\rm S_{\oplus}$\footnote{Retrieved from the NASA Exoplanet Archive on 2022 September 14, \url{https://exoplanetarchive.ipac.caltech.edu/cgi-bin/TblView/nph-tblView?app=ExoTbls&config=PSCompPars}}. The comparison sample contained 34 planets initially; we then removed the seven planets discovered via radial velocity as these are not known to transit. In Figure \ref{fig:TSM} we present this sample of small habitable zone and temperate exoplanets, highlighting the location of the conservative habitable zone \citep{2013ApJ...765..131K}. We also highlight the positions of the five TRAPPIST-1 planets that fall in this parameter space (d-h) \citep{trappist1}, the recently discovered LP 890-9\,c \citep[SPECULOOS-2\,c/TOI-4306\,c;][]{speculoos2}, K2-3\,d \citep{k2-3}, and the well characterised mini-Neptune, LHS 1140\,b \citep{lhs1140b}.

In the case of a `water world' composition, TOI-715\,b could have a transmission signal comparable to LHS-1140\,b if it does indeed have detectable atmospheric features. In the higher mass limit we find that TOI-715\,b has a TSM of approximately 1.9, below the suggested cutoff of 12 for follow-up of terrestrial planets.

We use \textsc{Tierra} \citep{tierra} to calculate simulated transmission spectra for TOI-715\,b in the `rocky' ($\sim 7\,\rm M_\oplus$) and `water-world' ($\sim 2\,\rm M_\oplus$) mass scenarios in the case of a primordial hydrogen/helium-dominated atmosphere. We then use \textsc{Pandexo} \citep{pandexo} to simulate \textit{JWST} observations in both cases to assess the detectability of atmospheric features. We find that in the low-mass case atmospheric features could be detectable with a single \jwst transit if the atmosphere is cloudless. In the higher mass case, we find that extracting meaningful features would require at least five transits, also in the cloud-free case. A secondary atmosphere with higher mean molecular mass could suppress the scale height by more than an order of magnitude, in turn making it challenging to detect atmospheric features of these planets \citep{2016Natur.537...69D}.

In both mass cases, the carbon dioxide feature centered on $ \rm 4.5\,\mu m$, the methane feature at $\rm 3.3\,\mu m$ and multiple water bands are some of the most accessible atmospheric features with the NIRSpec PRISM, the most suitable instrument for JWST follow-up given the similar brightness of TOI-715 compared to TRAPPIST-1. Detection of these features will constrain the metallicity of the putative planetary atmospheres, providing first hints of the carbon-chemistry of the atmosphere itself \citep{2012ApJ...758...36M}, and insights into their formation history \citep{2011ApJ...743L..16O}.

\subsection{Flares and habitability}
\label{sec:flares}

Both TOI-715\,b and the candidate TIC~271971130.02 lie in an interesting location of the parameter space with regard to insolation, radius, and potential rocky composition, naturally posing questions of their habitability. Many recent prebiotic chemistry and astrobiology studies investigated how life may have originated on Earth and other planets (see e.g. \citealt{Patel2015, Airapetian2016, Xu2018} and reviews by \citealt{Sutherland2017, Kitadai2018}). The first processes leading from inorganic compounds towards RNA precursors likely require a liquid solution (e.g. liquid surface water) and an energy source. For the latter, the host star's flaring may provide the necessary UV radiation (in the 200-280\,nm range) to trigger these early steps \cite[e.g.][]{Todd2018, Rimmer2018, Rimmer2021}.

On the other hand, stellar activity can also pose a significant danger to exoplanets. Strong stellar winds and XUV outbursts can contribute to atmospheric loss \citep[e.g.][]{Atri2020}. Even if the atmosphere prevails, coronal mass ejections (charged particle streams) may interact with and dissociate atmospheric ozone \citep[e.g.][]{Tilley2020}. Without the major atmospheric absorber of harmful UV radiation, the next flares might sterilize existing surface biology.

We investigate TOI-715's 24 sectors of TESS data for signs of stellar flaring. We find an average rate of 7 flares per 100 days of observations, with a maximum flare amplitude of 5\% in relative flux over nearly 2 years of data. This analysis was performed using the \textsc{stella} neural network \citep{Feinstein2020} and as part of the TESS flare catalog (G\"unther et al., in prep.), building on \citep{Guenther2020}.
We also compare this flare frequency distribution with the theoretical potential for ozone sterilisation \citep{Tilley2020} and laboratory thresholds for prebiotic chemistry \citep{Rimmer2018}. We find that the flaring is likely too rare and not energetic enough to influence either of these effects.

As TOI-715 is an older star (6.6$^{+3.2}_{-2.2}$~Gyr), its flaring is likely not as prominent as in younger years, when its planets were forming and prebiotic chemistry steps might have been triggered. Demographic studies of young (< 100 Myr) and older M dwarfs show that flaring activity decreases with stellar age, likely due to spin-down and a decreasing stellar dynamo \citep[e.g.][]{Guenther2020, Feinstein2020}. Thus, one could carefully speculate: if stronger flaring or other processes (volcanoes, impacts, or lightning) had led to astrobiology on TOI-715\,b or TIC~271971130.02 several Gyr ago, there might be a chance it could still exist (pending, of course, all other criteria such as atmospheric composition, liquid water, etc. are fulfilled).

\section{Conclusions}
\label{sec:conc}

In this work we have presented the discovery, validation and characterisation of TOI-715\,b, a $R_{\rm b}= \rm 1.550\pm0.064\,R_{\oplus}$ habitable zone planet orbiting an M4 star with a period of $19.288004_{-0.000024}^{+0.000027}$ days. We also demonstrated that there is possibly a second, smaller planet with radius $R_{02} = 1.066\pm0.092 ~\rm R_\oplus$ at a period of $25.60712_{-0.00036}^{+0.00031}$ days, placing it just inside the outer edge of the circumstellar habitable zone. This system represents the first \tess discovery to fall within this most conservative and widely applicable `habitable zone'.

We have also demonstrated that TOI-715\,b is amenable to further characterisation with precise radial velocities and transmission spectroscopy; detailed follow-up of this planet is crucial to further our understanding of the formation and evolution of small, close-in planets. Given its location in radius-instellation space, TOI-715\,b could also help us understand the characteristics of the M-dwarf radius valley. 

Confirmation of the existence of the TIC~271971130.02 in the coming months will also be crucial in planning further characterisation of planet b, as disentangling the two planets in a putative radial velocity campaign could prove challenging if the orbit of a second planet is not well known.

\section*{List of affiliations}
$^{1}$School of Physics \& Astronomy, University of Birmingham, Edgbaston, Birmingham B15 2TT, United Kingdom\\ 
$^{2}$Astrobiology Research Unit, University of Li\`ege, All\'ee du 6 ao\^ut, 19, 4000 Li\`ege (Sart-Tilman), Belgium\\ 
$^{3}$Dpto. Física Teórica y del Cosmos, Universidad de Granada, 18071, Granada, Spain\\
$^{4}$Center for Astrophysics and Space Sciences, UC San Diego, UCSD Mail Code 0424, 9500 Gilman Drive, La Jolla, CA 92093-0424, USA \\
$^{5}$Department of Earth, Atmospheric and Planetary Sciences, MIT, 77 Massachusetts Avenue, Cambridge, MA 02139, USA\\
$^{6}$Instituto de Astrof\'isica de Canarias (IAC), Calle V\'ia L\'actea s/n, 38200, La Laguna, Tenerife, Spain\\
$^{7}$Center for Astrophysics \textbar \ Harvard \& Smithsonian, 60 Garden St, Cambridge, MA 02138, USA\\
$^{8}$Univ. Grenoble Alpes, CNRS, IPAG, F-38000 Grenoble, France\\
$^{9}$Observatoire de Genève, Département d’Astronomie, Université de Genève, Chemin Pegasi 51b, 1290 Versoix, Switzerland \\
$^{10}$AIM, CEA, CNRS, Universit\'e Paris-Saclay, Universit\'e de Paris, F-91191 Gif-sur-Yvette, France\\ 
$^{11}$European Space Agency (ESA), European Space Research and Technology Centre (ESTEC), Keplerlaan 1, 2201 AZ Noordwijk, The Netherlands\\ 
$^{12}$NASA Ames Research Center, Moffett Field, CA 94035, USA\\ 
$^{13}$Department of Astrophysical and Planetary Sciences, University of Colorado Boulder, Boulder, CO 80309, USA\\ 
$^{14}$Department of Physics \& Astronomy, Vanderbilt University, 6301 Stevenson Center Ln., Nashville, TN 37235, USA \\
$^{15}$Department of Physics and Kavli Institute for Astrophysics and Space Research, Massachusetts Institute of Technology, Cambridge, MA 02139, USA\\ 
$^{16}$NASA Exoplanet Science Institute, Caltech/IPAC, Mail Code 100-22, 1200 E. California Blvd., Pasadena, CA 91125, USA\\ 
$^{17}$University of Bern, Center for Space and Habitability, Gesellschaftsstrasse 6, 3012 Bern, Switzerland\\
$^{18}$Department of Astronomy and Tsinghua Centre for Astrophysics, Tsinghua University, Beijing 100084, China\\ 
$^{19}$Instituto de Astronomía, Universidad Nacional Autónoma de México, Ciudad Universitaria, Ciudad de México, 04510, México\\ 
$^{20}$Cavendish Laboratory, JJ Thomson Avenue, Cambridge CB3 0HE, UK\\ 
$^{21}$Campo Catino Astronomical Observatory, Regione Lazio, Guarcino (FR), 03010 Italy\\  
$^{22}$Space Sciences, Technologies and Astrophysics Research (STAR) Institute, Universit\'e de Li\`ege, All\'ee du 6 Ao\^ut 19C, B-4000 Li\`ege, Belgium\\ 
$^{23}$Departamento de Astrof\'isica, Universidad de La Laguna (ULL), E-38206 La Laguna, Tenerife, Spain\\ 
$^{24}$Instituto de Astrof\'isica de Andaluc\'ia (IAA-CSIC), Glorieta de la Astronom\'ia s/n, 18008 Granada, Spain\\ 
$^{25}$Space Telescope Science Institute, 3700 San Martin Drive, Baltimore, MD, 21218, USA\\ 
$^{26}$Department of Astronomy, University of Maryland, College Park, MD  20742, USA\\ 
$^{27}$NASA Goddard Space Flight Center, 8800 Greenbelt Rd, Greenbelt, MD 20771, USA\\ 
$^{28}$Department of Aeronautics and Astronautics, Massachusetts Institute of Technology, Cambridge, MA 02139, USA\\ 
$^{29}$Kotizarovci Observatory, Sarsoni 90, 51216 Viskovo, Croatia\\ 
$^{30}$Hazelwood Observatory\\ 
$^{31}$SETI Institute, Mountain View, CA 94043, USA\\
$^{32}$Department of Astrophysical Sciences, Princeton University, Princeton, NJ 08544, USA\\ 

\section*{Acknowledgements}
We thank the reviewer for their comments and feedback as these helped clarify and streamline the maniscript significantly.
Funding for the TESS mission is provided by NASA's Science Mission Directorate. We acknowledge the use of public TESS data from pipelines at the TESS Science Office and at the TESS Science Processing Operations Center. This research has made use of the Exoplanet Follow-up Observation Program website, which is operated by the California Institute of Technology, under contract with the National Aeronautics and Space Administration under the Exoplanet Exploration Program. This paper includes data collected by the TESS mission that are publicly available from the Mikulski Archive for Space Telescopes (MAST).
Based on data collected by the SPECULOOS-South Observatory at the ESO Paranal Observatory in Chile.The ULiege's contribution to SPECULOOS has received funding from the European Research Council under the European Union's Seventh Framework Programme (FP/2007-2013) (grant Agreement n$^\circ$ 336480/SPECULOOS), from the Balzan Prize and Francqui Foundations, from the Belgian Scientific Research Foundation (F.R.S.-FNRS; grant n$^\circ$ T.0109.20), from the University of Liege, and from the ARC grant for Concerted Research Actions financed by the Wallonia-Brussels Federation. This work is supported by a grant from the Simons Foundation (PI Queloz, grant number 327127).
This research is in part funded by the European Union's Horizon 2020 research and innovation programme (grants agreements n$^{\circ}$ 803193/BEBOP), and from the Science and Technology Facilities Council (STFC; grant n$^\circ$ ST/S00193X/1, and ST/W000385/1).
The material is based upon work supported by NASA under award number 80GSFC21M0002
Based on data collected by the TRAPPIST-South telescope at the ESO La Silla Observatory. TRAPPIST is funded by the Belgian Fund for Scientific Research (Fond National de la Recherche Scientifique, FNRS) under the grant FRFC 2.5.594.09.F, with the participation of the Swiss National Science Fundation (SNF). 
Based on data collected under the ExTrA project at the ESO La Silla Paranal Observatory. ExTrA is a project of Institut de Planétologie et d'Astrophysique de Grenoble (IPAG/CNRS/UGA), funded by the European Research Council under the ERC Grant Agreement n. 337591-ExTrA. This work has been supported by a grant from Labex OSUG@2020 (Investissements d'avenir -- ANR10 LABX56).
This work has been carried out within the framework of the NCCR PlanetS supported by the Swiss National Science Foundation.
Some of the observations in the paper made use of the High-Resolution Imaging instrument Zorro obtained under Gemini LLP Proposal Number: GN/S-2021A-LP-105. Zorro was funded by the NASA Exoplanet Exploration Program and built at the NASA Ames Research Center by Steve B. Howell, Nic Scott, Elliott P. Horch, and Emmett Quigley. Zorro was mounted on the Gemini South telescope of the international Gemini Observatory, a program of NSF’s OIR Lab, which is managed by the Association of Universities for Research in Astronomy (AURA) under a cooperative agreement with the National Science Foundation. on behalf of the Gemini partnership: the National Science Foundation (United States), National Research Council (Canada), Agencia Nacional de Investigación y Desarrollo (Chile), Ministerio de Ciencia, Tecnología e Innovación (Argentina), Ministério da Ciência, Tecnologia, Inovações e Comunicações (Brazil), and Korea Astronomy and Space Science Institute (Republic of Korea).
The Digitized Sky Surveys were produced at the Space Telescope Science Institute under U.S. Government grant NAG W-2166. The images of these surveys are based on photographic data obtained using the Oschin Schmidt Telescope on Palomar Mountain and the UK Schmidt Telescope. The plates were processed into the present compressed digital form with the permission of these institutions.
This work makes use of observations from the LCOGT network. Part of the LCOGT telescope time was granted by NOIRLab through the Mid-Scale Innovations Program (MSIP). MSIP is funded by NSF.
This publication benefits from the support of the French Community of Belgium in the context of the FRIA Doctoral Grant awarded to MT.
MNG acknowledges support from the European Space Agency (ESA) as an ESA Research Fellow. 
BVR thanks the Heising-Simons Foundation for support.
MG is F.R.S.-FBRS Research Director. 
YGMC acknowledges support from UNAM-PAPIIT-IG101321
FJP acknowledges financial support from the grant CEX2021-001131-S funded by MCIN/AEI/10.13039/501100011033.
This research made use of Lightkurve, a Python package for Kepler and TESS data analysis (Lightkurve Collaboration, 2018). 
This work made use of Astropy:\footnote{http://www.astropy.org} a community-developed core Python package and an ecosystem of tools and resources for astronomy \citep{astropy:2013, astropy:2018, astropy:2022}. 
This research has made use of the NASA Exoplanet Archive, which is operated by the California Institute of Technology, under contract with the National Aeronautics and Space Administration under the Exoplanet Exploration Program. 
This research made use of \textsf{exoplanet} \citep{exoplanet} and its
dependencies \citep{exoplanet:agol20, exoplanet:arviz, exoplanet:astropy13,
exoplanet:astropy18, exoplanet:luger18, exoplanet:pymc3, exoplanet:theano}.
Resources supporting this work were provided by the NASA High-End Computing (HEC) Program through the NASA Advanced Supercomputing (NAS) Division at Ames Research Center for the production of the SPOC data products.

\section*{Data Availability}

\tess data products are available via the MAST portal at \url{https://mast.stsci.edu/portal/Mashup/Clients/Mast/Portal.html}. Follow-up photometry and high resolution imaging data for TOI-715 are available on ExoFOP at \url{https://exofop.ipac.caltech.edu/tess/target.php?id=271971130}. These data are freely accessible to ExoFOP members immediately and are publicly available following a one-year proprietary period.



\bibliographystyle{mnras}
\bibliography{TOI715} 








\bsp	
\label{lastpage}
\end{document}